%% file: main.tex
\documentclass[twocolumn]{aastex631}
\usepackage{booktabs}
\usepackage{multirow}
\usepackage{graphicx}

\providecommand{\teff}{\ensuremath{T_{\rm eff}}} 
\providecommand{\mj}{\ensuremath{\,{\rm M_{Jup}}}} 
\providecommand{\rj}{\ensuremath{\,{\rm R_{Jup}}}} 
\providecommand{\angstrom}{\(\text{\r{A}}\) } 

\shorttitle{Two Transiting Brown Dwarfs from TESS in the BD Desert}
\shortauthors{Zhang et al.}

\begin{document}

\title{An Oasis in the Brown Dwarf Desert: Confirmation of Two Low-mass Transiting Brown Dwarfs Discovered by TESS}

\author[0009-0004-0455-2424]{Elina Y. Zhang} 
\affiliation{Institute for Astronomy, University of Hawaii, 2680 Woodlawn Drive, Honolulu, HI 96822 USA}

\author[0000-0001-6416-1274]{Theron W. Carmichael}
\affiliation{Institute for Astronomy, University of Hawaii, 2680 Woodlawn Drive, Honolulu, HI 96822 USA}

\author[0000-0001-8832-4488]{Daniel Huber}
\affiliation{Institute for Astronomy, University of Hawaii, 2680 Woodlawn Drive, Honolulu, HI 96822 USA}

\author[0000-0002-3481-9052]{Keivan G. Stassun}
\affiliation{Department of Physics and Astronomy, Vanderbilt University, Nashville, TN 37235, USA}

\author[0000-0002-4909-5763]{Akihiko Fukui}
\affiliation{Komaba Institute for Science, The University of Tokyo, 3-8-1 Komaba, Meguro, Tokyo 153-8902, Japan}
\affiliation{Instituto de Astrof\'{i}sica de Canarias (IAC), E-38205 La Laguna, Tenerife, Spain}

\author[0000-0001-8511-2981]{Norio Narita}
\affiliation{Komaba Institute for Science, The University of Tokyo, 3-8-1 Komaba, Meguro, Tokyo 153-8902, Japan}
\affiliation{Instituto de Astrof\'{i}sica de Canarias (IAC), E-38205 La Laguna, Tenerife, Spain}
\affiliation{Astrobiology Center, 2-21-1 Osawa, Mitaka, Tokyo 181-8588, Japan}

\author[0000-0001-9087-1245]{Felipe Murgas}
\affiliation{Instituto de Astrof\'{i}sica de Canarias (IAC), E-38205 La Laguna, Tenerife, Spain}
\affiliation{Departamento de Astrof\'{i}sica, Universidad de La Laguna (ULL), E-38206 La Laguna, Tenerife, Spain}

\author[0000-0003-0987-1593]{Enric Palle}
\affiliation{Instituto de Astrof\'{i}sica de Canarias (IAC), E-38205 La Laguna, Tenerife, Spain}
\affiliation{Departamento de Astrof\'{i}sica, Universidad de La Laguna (ULL), E-38206 La Laguna, Tenerife, Spain}

\author[0000-0001-9911-7388]{David W. Latham}
\affiliation{Center for Astrophysics, Harvard \& Smithsonian, 60 Garden Street, Cambridge, MA 02138, USA}

\author[0000-0002-2830-5661]{Michael L. Calkins}
\affiliation{Center for Astrophysics, Harvard \& Smithsonian, 60 Garden Street, Cambridge, MA 02138, USA}

\author[0000-0002-6892-6948]{Sara Seager}
\affiliation{Department of Physics and Kavli Institute for Astrophysics and Space Research, Massachusetts Institute of Technology, Cambridge, MA 02139, USA}
\affiliation{Department of Earth, Atmospheric and Planetary Sciences, Massachusetts Institute of Technology, Cambridge, MA 02139, USA}
\affiliation{Department of Aeronautics and Astronautics, MIT, 77 Massachusetts Avenue, Cambridge, MA 02139, USA}

\author[0000-0002-4265-047X]{Joshua N. Winn}
\affiliation{Department of Astrophysical Sciences, Princeton University, Princeton, NJ 08544, USA}

\author{Michael Vezie}
\affiliation{Department of Physics and Kavli Institute for Astrophysics and Space Research, Massachusetts Institute of Technology, Cambridge, MA 02139, USA}

\author[0000-0002-0476-4206]{Rebekah Hounsell}
\affiliation{NASA Goddard Space Flight Center, 8800 Greenbelt Rd, Greenbelt, MD 20771, USA}

\author[0000-0002-4047-4724]{Hugh P. Osborn}
\affiliation{NCCR/Planet-S, Physikalisches Institut, Universität Bern, Gesellschaftsstrasse 6, 3012 Bern, Switzerland}

\author[0000-0003-1963-9616]{Douglas A. Caldwell}
\affiliation{SETI Institute, Mountain View, CA 94043 USA/NASA Ames Research Center, Moffett Field, CA 94035 USA}

\author[0000-0002-4715-9460]{Jon M. Jenkins}
\affiliation{NASA Ames Research Center, Moffett Field, CA 94035, USA}

\begin{abstract}
As the intermediate-mass siblings of stars and planets, brown dwarfs (BDs) are vital to study for a better understanding of how objects change across the planet-to-star mass range. Here, we report two low-mass transiting BD systems discovered by TESS, TOI-4776 (TIC 196286578) and TOI-5422 (TIC 80611440), located in an under-populated region of the BD mass--period space. These two systems have comparable masses but different ages. The younger and larger BD is TOI-4776b with $32.0^{+1.9}_{-1.8}\mj$ and $1.018^{+0.048}_{-0.043}\rj$, orbiting a late-F star about $5.4^{+2.8}_{-2.2}$ Gyr old in a 10.4138$\pm$0.000014 day period. The older TOI-5422b has $27.7^{+1.4}_{-1.1}\mj$ and $0.815^{+0.031}_{-0.026}\rj$ in a 5.3772$\pm$0.00001 day orbit around a subgiant star about $8.2\pm2.4$ Gyr old. Compared with substellar mass--radius (M--R) evolution models, TOI-4776b has an inflated radii. In contrast, TOI-5422b is slightly ``underluminous" with respect to model predictions, which is not commonly seen in the BD population. In addition, TOI-5422 shows apparent photometric modulations with a rotation period of 10.75$\pm$0.54 day found by rotation analysis, and the stellar inclination angle is obtained to be $I_{\star}=75.52^{+9.96}_{-11.79}$$^{\circ}$. Therefore, it is likely that TOI-5422b is spinning up the host star and its orbit is aligned with the stellar spin axis.
\end{abstract}

\keywords{Brown dwarfs; Transit photometry; Radial velocity; Substellar companion stars; Spectroscopy; Photometry}

\section{Introduction} \label{sec:intro}

Brown dwarfs (BDs) are defined by mass to be in between low-mass stars and giant planets. The dividing lines between these populations are 11-16$\mj$ \citep[deuterium fusion limit,][]{Spiegel2011} and 75-80$\mj$ \citep[hydrogen fusion limit,][]{Baraffe2002, Chabrier2023}. As the intermediate-mass siblings of stars and planets, BDs are vital to study for an understanding of the initial mass function, star and planet formation, binary system evolution, and atmosphere of giant planets.

How brown dwarfs are formed is still under debate. Two popular theories are star-like formation in molecular clouds via gravitational instability \citep{PadoanNordlund2004, Hennebelle2008, Kratter2016} and planet-like formation in protoplanetary disks through core accretion \citep{Alibert2005, Mordasini2009}. Core accretion is thought to be the dominant mechanisms for less-massive objects up to 4-10$\mj$ \citep{Schlaufman2018}, and gravitational instability favors forming more massive objects. However, there is a likely overlap in the mass ranges for each mechanism. The most massive object with strong evidence to form via core accretion is CWW 89Ab, with 36.5$\mj$, and it represents the very tail end of the expected mass of the core accretion formation pathway \citep{Beatty2018}.

Another open question in brown dwarf science is the accuracy of evolutionary models. In particular, the radius of a BD is expected to contract as it ages, since they lack a sustaining mechanism to initiate a mechanism like hydrostatic equilibrium seen in stars. The contraction will be quicker at young ages, and then asymptotically decelerates \citep{Phillips2020}. Therefore, the radius for a given BD mass will reach a plateau at ages older than a few Gyr. It is worth to mention that BD will first fuse deuterium in the core at very young ages, causing a temporary increase in radius, but then gradually cool and contract as deuterium in the core exhausts quickly \citep{Burrows2011}. BDs are comparable to the size of Jupiter, with a radius range of roughly 0.8$\rj$ to 1.5$\rj$ for BDs transiting main sequence stars (older than 100 Myr). For BD candidates transiting pre-main sequence host stars or particularly young stars less than 100 Myr, larger radii up to 3-5$\rj$ are expected and even observed \citep[e.g. RIK 72b;][]{RIK72b}. Several known BDs have radii that lie far from modeled evolution tracks. A handful of BDs are found to have inflated radius, and they expand the entire mass range from low to high. For the cases of KELT-1b \citep{KELT-1b}, GPX-1b \citep{GPX-1b}, and TOI-2336b \citep{TOI-2336b}, the authors argue that irradiation is the cause of inflation. While for ZTF J2020+5033 \citep{ZTFJ2020+5033} and KOI-189b \citep{KOI-189b} the argument is that magnetic braking within the brown dwarf might cause the inflation. On the contrary, BDs that is smaller in radius compared to models \citep[e.g. TOI-569b;][]{Carmichael2020} may be described as ``underluminous", although we don't technically measure its luminosity. Considering the known BDs that show discrepancies between their radii and model predictions, we need more BD case studies to understand why the discrepancy exists and to improve our substellar evolution models.

Transiting BDs provide valuable information on the radius from transit photometry. The radial velocity (RV) method alone provides only a lower mass limit, but when combined with information from transit photometry, this minimum mass $M\sin{i}$ degeneracy can be broken. The TESS (Transiting Exoplanet Survey Satellite) mission can determine the orbital inclination $i$ from light curves. However, determining the age of BDs is much more difficult than finding their mass and radius. Relatively accurate age estimation can be obtained by measuring the age of sun-like stars the BD transits, as stellar ages are more well-modeled by fitting stellar isochrones. Combining the mass, radius, and age information, transiting BDs are clearly in an advantageous position to test substellar evolutionary models. 

Previous studies have explored tidal effects between the host star and BD. These tidal interactions can help us understand different formation pathways that may result in the system's current configuration. For instance, \cite{Mazeh2008} and \cite{Grieves2017} proposed a transition from lower-mass BDs with near circular orbits to more massive BDs with more diverse eccentricity at $\sim 42.5 \mj$ based on a mass threshold proposed by \cite{MaGe2014}. \cite{Bowler2020} found that systems with high mass ratios ($M_{BD}/M_{\star}>0.01$) have a broader eccentricity distribution, so they are more likely to have high eccentricities, while low-mass-ratio systems ($<$0.01) tend to have lower eccentricities. So far there are only six transiting BD systems that have obliquity measured through the Rossiter-McLaughlin observations: CoRoT-3b \citep{CoRoT-3b}, KELT-1b \citep{KELT-1b}, WASP-30b \citep{WASP-30b}, HATS-70b \citep{HATS-70b}, and recently TOI-2533b \citep{TOI-2533b} and GPX-1b \citep{GPX-1bRM}. Although the sample size is small, it hints that unlike planet companions that can have a diverse range in orbital obliquity, BDs tend to be more aligned with their host stars.

The TESS mission has been responsible for the vast majority of new transiting BDs, granting us new perspectives on these objects. Up to now, there are nearly 50 transiting BD system discovered, with more that 2/3 of them were discovered by TESS. Although the current known transiting BD sample is still scant, it is rapidly growing. Combining data from NASA’s TESS mission, the Gaia mission, and ground-based follow-up observations, astronomers diligently search for and characterize new transiting BDs. Among known transiting BDs, the mass-radius (M-R) relationship shows an apparent lack of objects at $\sim 30 \mj$, coincident with the “Brown Dwarf Desert”, which is the relatively low occurrence of short-period BD companions to main-sequence host stars compared with planetary and stellar companions \citep{GretherLineweaver2006}. Therefore, adding transiting BDs to the sample with masses in the range corresponding to the desert is desirable.

In this paper, we present the discovery and characterization of two transiting BD systems, TOI-4776 (TIC 196286578) and TOI-5422 (TIC 80611440), discovered by TESS. Both of them are located in a relatively sparsely populated region in substellar mass-radius space. For each system, we did a global analysis on the light curve, radial velocity, and stellar SED to derive the host star and BD properties. We measured the radii and masses of both brown dwarfs, and used constraints on the host star's age to test evolutionary models. These two targets have comparable masses but different ages. Current substellar mass–radius evolutionary models will predict that TOI-4776b represents younger BDs with rapid radius contraction, while TOI-5422b is an older BD representative whose radius contraction slows to a constant value. 

\section{Observations} \label{sec:observation}

Transits of TOI-5422b were detected by MIT’s Quick Look Pipeline \citep[QLP;][]{Huang2020a, Huang2020b} and contemporaneously by the SPOC in the TESS-SPOC FFI light curve \citep{Caldwell2020} in transit searches of Sector 45. The TESS Science Office (TSO) reviewed the QLP vetting reports and issued an alert on 20 April 2022 \citep{Guerrero2021}. The difference image centroiding test \citep{Twicken2018} located the host star within 1.9$\pm$2.5 arcsec of the transit source. This constraint was tightened to 0.869$\pm$2.6 arcsec in the search of the 2-min data for sectors 71 \& 72.
 
TOI-4776b was detected by the FAINT search pipeline \citep{Kunimoto2021} in a search of sectors 7, 8 \& 34. The TSO reviewed the vetting report and issued a TOI alert on 21 December 2021. The SPOC detected the transit signature subsequently in a search of sectors 71 \& 72 and constrained the location of the host star to within 0.895$\pm$2.6 arcsec of the transit source.

All the TESS data used in this paper can be found in MAST: \dataset[10.17909/v9bb-w005]{http://dx.doi.org/10.17909/v9bb-w005}.

\subsection{TESS and Ground-based Light Curves} 
\subsubsection{TESS Light Curves for TOI-5422} 
\label{sec:TESS_lc5422}
\begin{figure*}[!htb]
    \centering
    \includegraphics[width=1.0\textwidth]{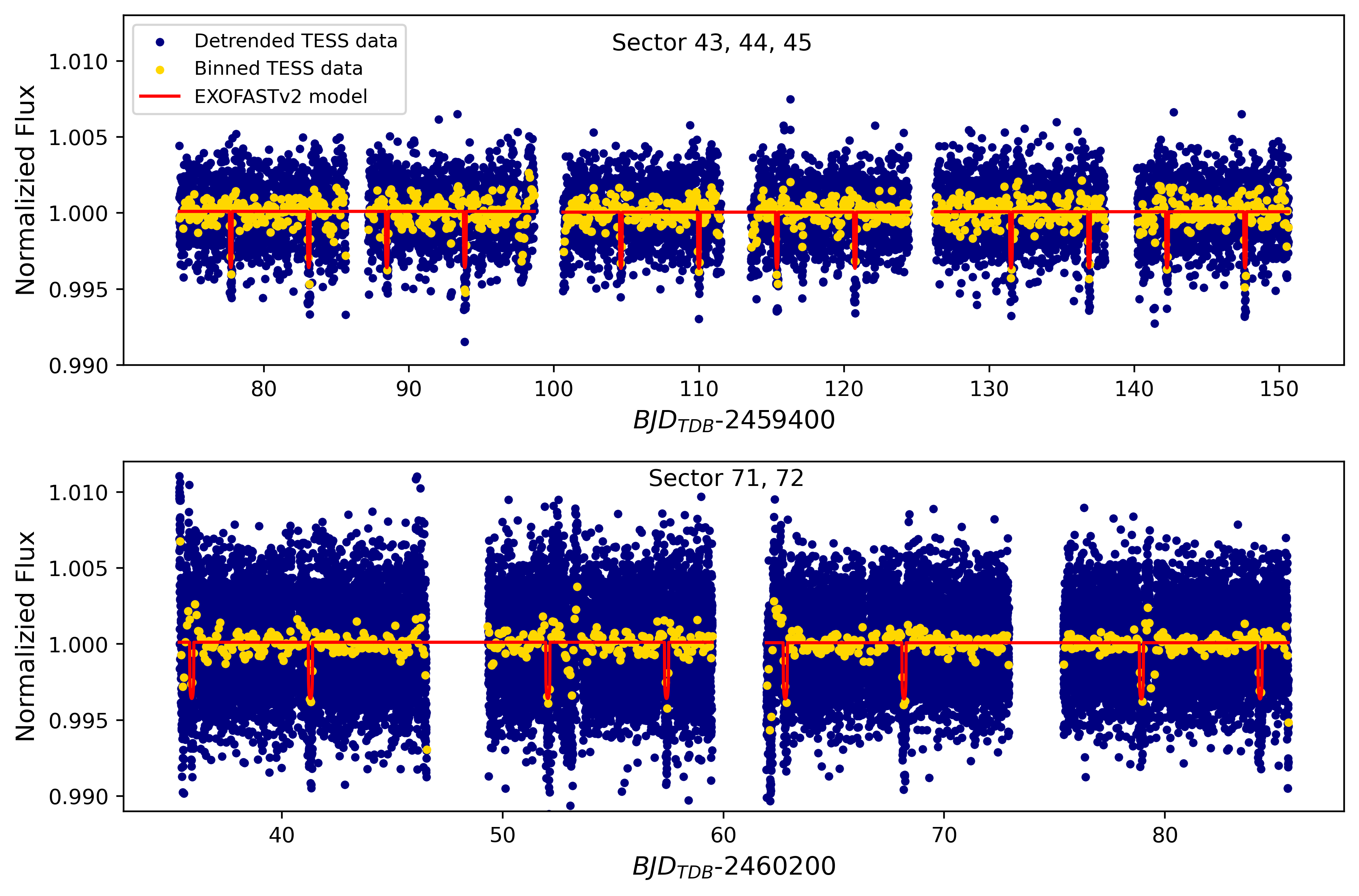}
    \vspace{-0.9em}
    \caption{\normalsize Detrended TESS light curve of TOI-5422 in the dark blue points. The star was observed at 10 minutes cadence in TESS sectors 43, 44, and 45, and 2 minutes cadence in sectors 71 and 72. The binning shown here as yellow points uses bin sizes of 90 minutes. This star also exhibits photometric variations likely due to stellar rotation; these effects have been removed for the transit analysis. The red line is the fitted model from {\fontfamily{qcr}\selectfont EXOFASTv2}.}
\label{fig:5422_transit}
\end{figure*}

TOI-5422 was observed by the TESS mission in sectors 43, 44, and 45 at a 10-minute cadence, and sectors 71 and 72 at a 2-minute cadence. We used light curves produced by the MIT Quick Look Pipeline \citep[QLP;][]{Huang2020a, Huang2020b} for sectors 43-45, and the Presearch Data Conditioning Simple Aperture Photometry flux \citep[PDCSAP;][]{Stumpe2014, Stumpe2012, Smith2012} for sectors 71 and 72 produced by the Science Processing Operations Center \citep[SPOC;][]{Jenkins2016}. These data are available at the Mikulski Archive for Space Telescopes (\href{https://mast.stsci.edu/portal/Mashup/Clients/Mast/Portal.html}{MAST}). We normalized and flattened the light curves with the {\fontfamily{qcr}\selectfont lightkurve} package in Python \citep {lightkurve2018}. We used {\fontfamily{qcr}\selectfont lightkurve} package's built-in flattening tool, which removes the low frequency trend due to stellar and instrumental variability as well as scattered background light. We used lengths of the filter window (i.e. the number of coefficients) range from 301 to 341 for sector 43-45 and 1551 for sector 71 and 72 in the flattening tool to remove long-term trends. We removed the large spikes and dips around the data gaps between sector 43-45. The detrended TESS light curves for TOI-5422 are shown in \textbf{Figure \ref{fig:5422_transit}}. 

\begin{figure}[!hbt]
    \centering
    \includegraphics[width=\columnwidth]{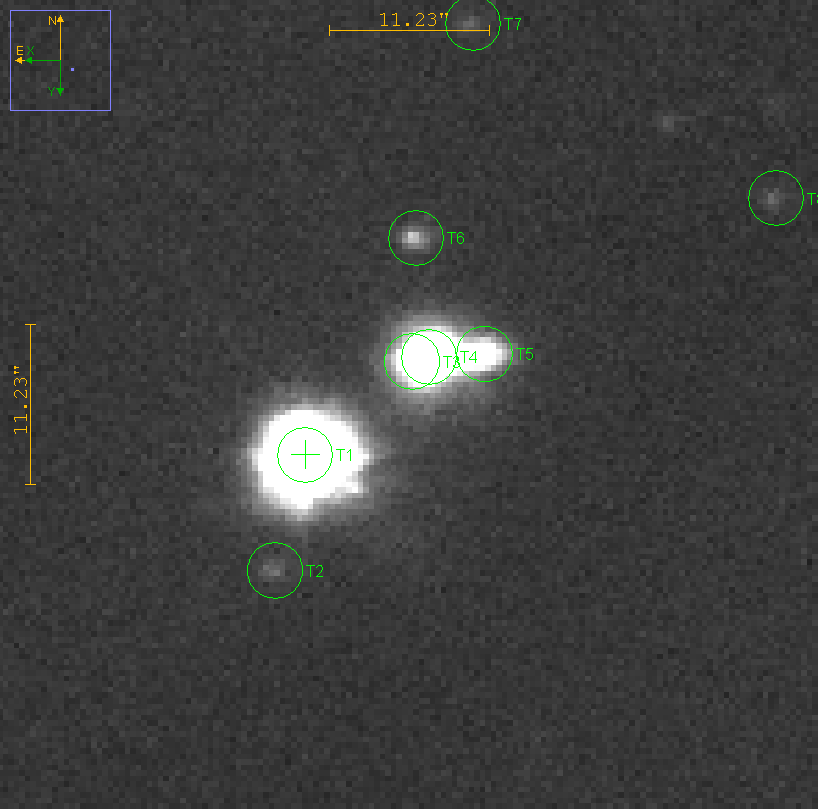}
    \caption{\normalsize The zoomed field of TOI-4776 when LCO-CTIO made the ground-based followup. T4 is our target TOI-4776, and T1 is TOI-592.}
\label{fig:4776_field}
\end{figure}

\begin{figure} [!hbt]
    \centering
    \includegraphics[width=\columnwidth]{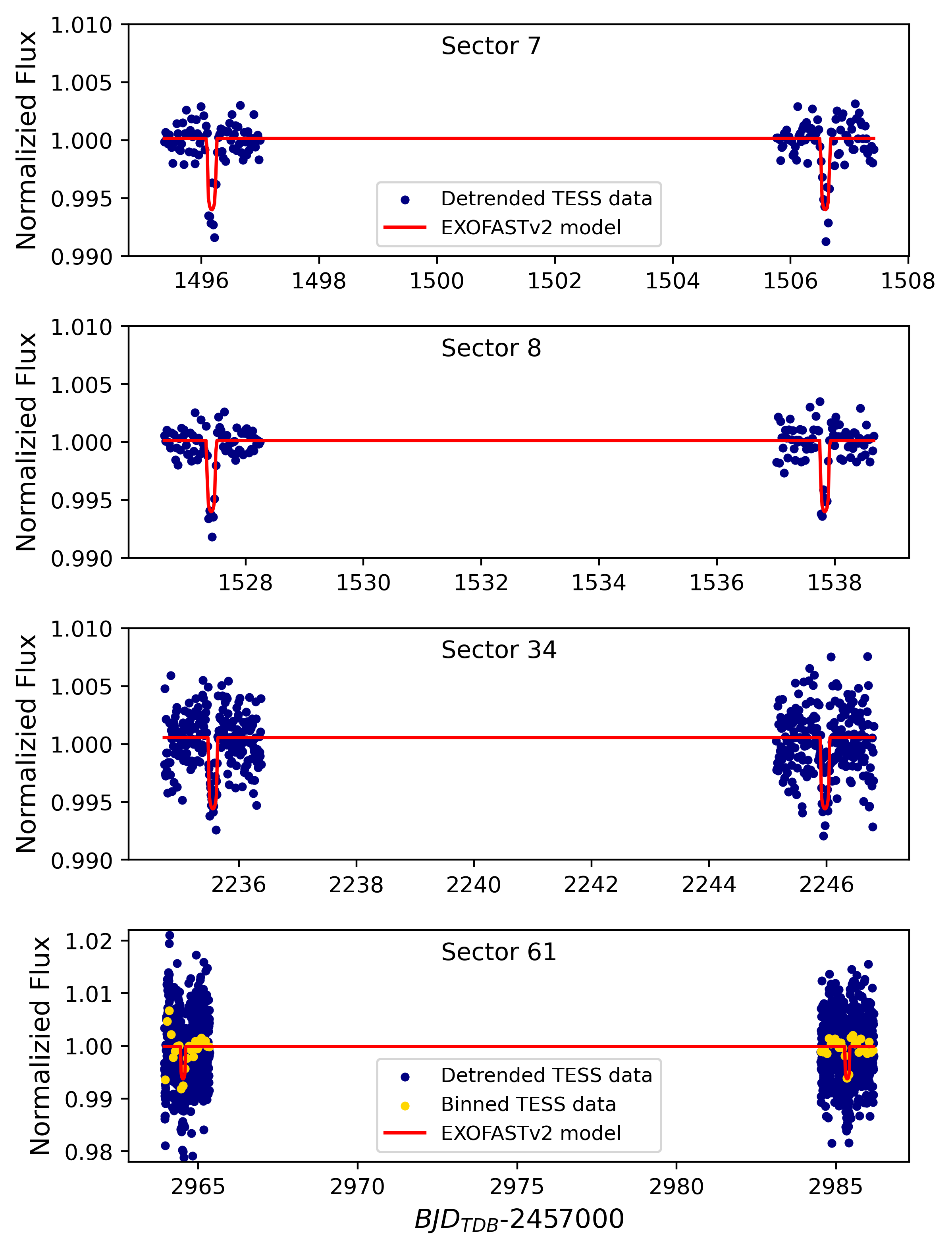}
    \vspace{-0.9em}
    \caption{\normalsize Detrended TESS light curve of TOI-4776 in the dark blue points. The star was observed at 30 minutes cadence in TESS sectors 7 \& 8, 10 minutes cadence in sector 34, and 200-second cadence in sector 61.
    The binning shown in sector 61 as yellow points uses bin sizes of 90 minutes. The red line is the fitted model from {\fontfamily{qcr}\selectfont EXOFASTv2}. The data for all 4 sectors center around the mid-transit time with a length of 10 times the transit duration.}
\label{fig:4776_TESS_transit}
\end{figure}

\begin{figure} [!hbt]
    \centering
    \includegraphics[width=\columnwidth]{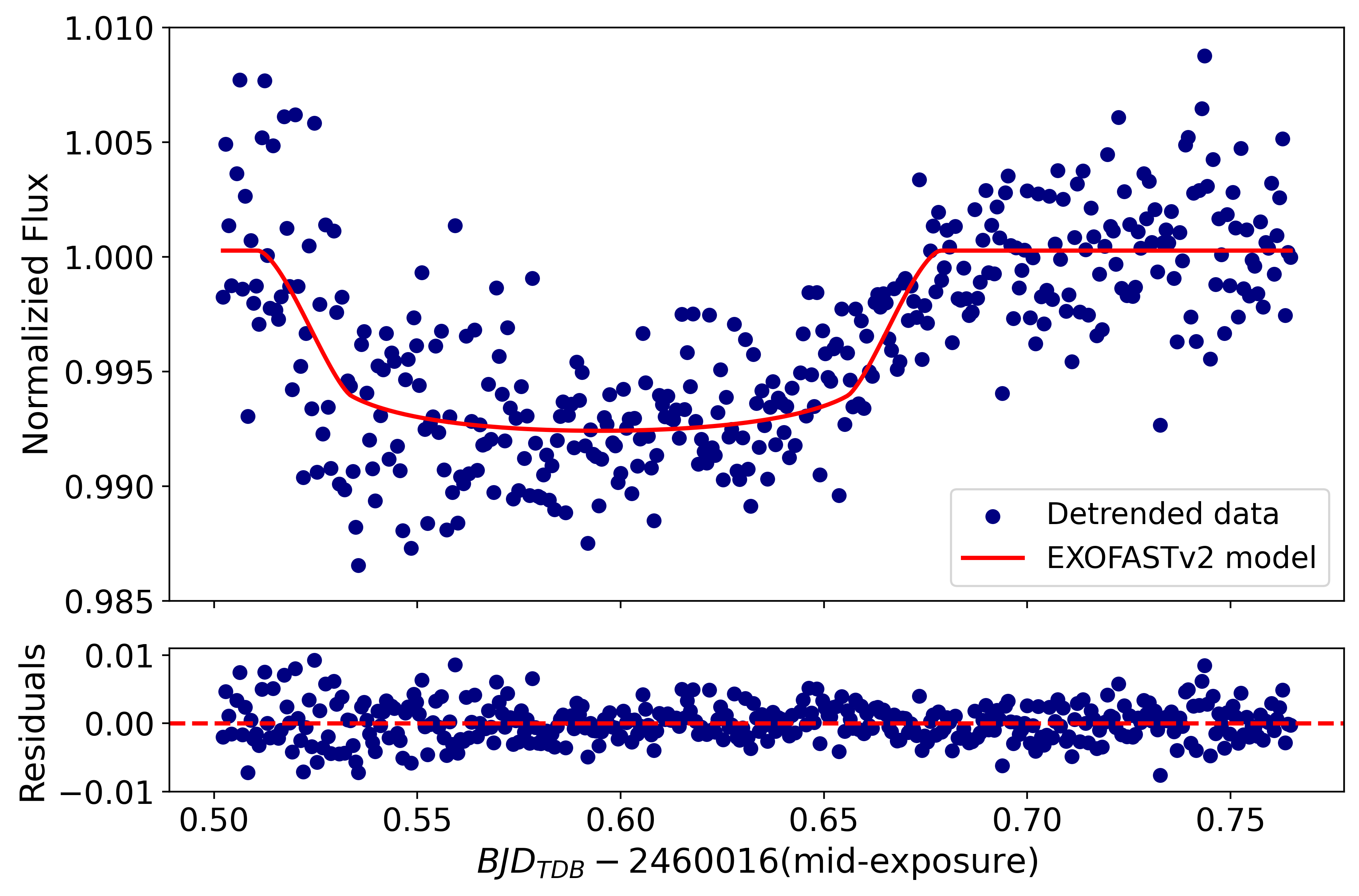}
    \vspace{-0.9em}
    \caption{\normalsize Detrended LCO-CTIO light curve with residuals of TOI-4776 in the dark blue points. The red line is the fitted model from {\fontfamily{qcr}\selectfont EXOFASTv2}.}
\label{fig:4776_transit}
\end{figure}

\subsubsection{MuSCAT2 Observations for TOI-5422}
TOI-5422 was observed on 2024 February 2 using the multi-band imager MuSCAT2 \citep{Narita2019}, mounted on the 1.5 m Telescopio Carlos S\'{a}nchez (TCS) at the Teide Observatory, Spain. MuSCAT2 is equipped with four CCDs, allowing for simultaneous imaging in the $g'$, $r'$, $i'$, and $z_s$ bands with short readout times. Each CCD consists of 1024 $\times$ 1024 pixels, covering a field of view of $7.4 \times 7.4$ arcmin$^2$.

To avoid saturation of the target star, the telescope was deliberately defocused during the observations. The exposure times were set to 20, 15, 20 and 20 seconds for the $g'$, $r'$, $i'$ and $z_s$ bands respectively. The raw data were processed through the MuSCAT2 pipeline \citep{Parviainen2019}, which performs standard reductions such as dark and flat field calibrations, aperture photometry, and transit model fitting, taking into account the instrumental systematics. However, the data were partially affected by clouds, and the egress of the transit could not be detected on the target star with high significance. Nevertheless, the observations were instrumental in ruling out potential eclipsing binaries (EBs) that could mimic the transit signal, eliminating 38 stars within 2.5 arcseconds of the target as possible EBs.

This MuSCAT2 observation helped us confirm that the transit of TOI-5422b is on the correct target, but it is not high enough SNR for fitting. Therefore, we didn't include the data in our analysis.

\subsubsection{TESS Light Curve for TOI-4776}
\label{sec:TESS_lc4776}
TOI-4776 was observed by TESS in sector 7 and 8 with a 30-minute cadence, in sector 34 with a 10-minute cadence, and in sector 61 with a 200-second cadence. However, another star, TOI-592, is present within the same TESS pixel (10.94'' separation), along with several other fainter stars at even closer separations (see \textbf{Figure \ref{fig:4776_field}}), making it more difficult to do the analysis. For instance, the CROWDSAP metric for TESS sector 61 is 0.30545211 and FLFRCSAP=0.66195452, which means that only 30.5\% of the light in the aperture is from the target star, so 69.5\% of the flux is from other nearby stars, and 66\% of the light in the target star’s PSF is captured in the photometric aperture. Therefore, we need to correct the effects from the nearby star on our target, which makes the transit depths of TOI-4776 appear shallower. We applied a dilution factor on our TESS transit fitting, which is calculated using $A_D=F2/(F1 + F2)$, where F2 is the flux of the contaminating star (TOI-592) and F1 is the flux of our target. Moreover, the raw TESS light curves have large systematics, probably due to momentum dumps. Thus, before flattening the the data, we cut the out-of-transit part of the light curves and left with the data that center around the mid-transit with a length of 10 times the transit duration. In other words, we only used the data that is within 5 times the transit duration before or after the mid-transit. The detrended TESS light curve of TOI-4776 is shown in \textbf{Figure \ref{fig:4776_TESS_transit}}.

\subsubsection{LCO-CTIO 1m Light Curve for TOI-4776} \label{sec:ground_lc4776}
In addition to the TESS light curves for TOI-4776, we also used its ground-based photometric follow-up by the Las Cumbres Observatory (LCOGT; \cite{Brown2023}). The transit was observed on 2023 March 12 using 30s exposures in Pan-STARRS z-short band from the LCOGT 1.0 m node at Cerro Tololo Inter-American Observatory (CTIO). The 4096$\times$4096 LCOGT SINISTRO cameras have an image scale of 0.39 arc-seconds/pixel. The observation duration is 378.1 min. The images were calibrated by the standard LCOGT {\fontfamily{qcr}\selectfont BANZAI} pipeline, and photometric data were extracted with {\fontfamily{qcr}\selectfont AstroImageJ} with an average aperture radius of 1.9 arc-seconds. The detrended LCO-CTIO light curve of TOI-4776 is shown in \textbf{Figure \ref{fig:4776_transit}}.

\subsubsection{Source Mislabeling on Ground-based Followup for TOI-4776}
As shown in \textbf{Figure \ref{fig:4776_field}}, the zoomed field of TOI-4776 when LCO-CTIO made the ground-based followup is very crowded. Specifically, our target TOI-4776 (T4) and the brighter nearby star TOI-592 (T1) are only 10.94'' apart from each other. Therefore, when taking the transit data, there was a mismatch on which star the transit belonged to. The transit data on TOI-4776 was not propagated to the follow up observations page for TOI-4776 on the ExoFOP website \citep{EXOFOPweb, EXOFOPtess}, but instead was recorded under the name of \href{https://exofop.ipac.caltech.edu/tess/target.php?id=196286587}{TOI-592}. After checking the observation notes and radial velocity data on both stars, we confirmed that the transit was originating from TOI-4776 (T4), not TOI-592 as it appears on ExoFOP.

\subsection{TRES Spectra} \label{sec:RVdata}
We use the TRES (Tillinghast Reflector Echelle Spectrograph) instrument on Mt. Hopkins, Arizona to obtain spectra for both TOI-4776 and TOI-5422. TRES is a fiber-fed optical echelle spectrograph on the 1.5-meter Tillinghast telescope at the Smithsonian Astrophysical Observatory's Fred L. Whipple Observatory. It covers a wavelength range of 3900\angstrom to 9100\angstrom with a resolving power of $R\sim 44\,000$. We use multiple echelle orders for each spectrum to measure a relative radial velocity (RV) at each phase in the orbit of each of the transiting BDs. Each order is cross-correlated with the highest observed signal-to-noise (S/N) spectrum of the target star and then the average RV of all the orders per spectrum is taken as the RV of the star for that observation.

We use the stellar parameter classification (SPC) software package by \cite{Buchhave2012} to derive $\teff$, metallicity [Fe/H], $\log{g}$, and the projected stellar equatorial velocity $v\sin{I_\star}$ from the TRES spectra of TOI-4776 and TOI-5422. SPC uses a library of calculated spectra in the 5030-5320\angstrom wavelength range, centered near the Mg b triplet. For TOI-4776, we took a series of 9 spectra to characterize the orbit of this BD from 2022 January 7 to 2022 February 25 with a range of exposure times from 1600\,s and 3240\,s, giving us a S/N between 18 and 22, respectively (See \textbf{Table \ref{tab:4776_rvs}}). Using SPC, we derive the following stellar parameters for TOI-4776: $\teff  = 6008 \pm 34$K, $\log{g} = 4.27 \pm 0.07$, $\rm [Fe/H] = -0.06 \pm 0.06$, and $v\sin{I_\star}= 6.5 \pm 0.3$ $\rm km\, s^{-1}$. For TOI-5422, we took a series of 11 spectra to derive an orbital solution for the system from 2022 September 11 to 2022 October 05. The exposure times for these spectra range from 750\,s to 1300\,s to give a S/N range of 14 to 22 (See \textbf{Table \ref{tab:5422_rvs}}). The stellar parameters for these spectra derived with SPC for TOI-5422 are: $\teff = 5736 \pm 40$K, $\log{g} = 4.25 \pm 0.07$, $\rm [Fe/H] = -0.05 \pm 0.05$, and $v\sin{I_\star}= 8.1 \pm 0.2$ $\rm km\, s^{-1}$.

\begin{deluxetable}{ccccc}
\tablecaption{RVs from TRES for TOI-4776 \label{tab:4776_rvs}}
\tablehead{
\colhead{$\rm BJD_{\rm TDB}$} & \colhead{RV (m/s)} & \colhead{$\sigma_{\rm RV}$ (m/s)} & \colhead{S/N} & \colhead{Expo. (s)}
}
\startdata
2459586.865161 &   5428.7 &  44.1 &  19.2  & 1600 \\
2459622.805127 &   -12.5  &  32.2 &  20.3  & 2160 \\
2459623.798567 &   -85.4  &  30.8 &  22.1  & 3240 \\
2459624.704426 &   716.0  &  43.7 &  18.7  & 1900 \\
2459625.734900 &   2089.9 &  36.0 &  21.1  & 2200 \\
2459626.731843 &   3700.8 &  25.3 &  20.3  & 2300 \\
2459630.769817 &   3445.2 &  28.0 &  22.1  & 2200 \\
2459631.765406 &   1711.7 &  37.3 &  19.0  & 2160 \\
2459635.733391 &   1137.8 &  35.2 &  18.1  & 2000 \\
\enddata
\end{deluxetable}

\begin{deluxetable}{ccccc}
\tablecaption{RVs from TRES for TOI-5422 \label{tab:5422_rvs}}
\tablehead{
\colhead{$\rm BJD_{\rm TDB}$} & \colhead{RV (m/s)} & \colhead{$\sigma_{\rm RV}$ (m/s)} & \colhead{S/N} & \colhead{Expo. (s)}
}
\startdata
2459834.003684 &   -505.0 &  35.7 &  17.8  & 750  \\
2459836.982013 &   5546.9 &  32.3 &  16.9  & 900  \\
2459837.998346 &   2516.6 &  33.6 &  20.2  & 1140 \\
2459838.964136 &   -77.1  &  36.4 &  22.1  & 1300 \\
2459842.008145 &   5460.0 &  35.5 &  19.1  & 1080 \\
2459850.012876 &   -375.9 &  45.0 &  14.4  & 1100 \\
2459853.943687 &   3287.9 &  38.4 &  18.2  & 900  \\
2459855.004195 &   26.6   &  33.2 &  17.6  & 800  \\
2459856.003290 &   -314.9 &  38.5 &  17.3  & 875  \\
2459856.894852 &   1681.2 &  29.6 &  18.0  & 850  \\
2459857.906581 &   5153.4 &  30.6 &  18.4  & 960  \\
\enddata
\end{deluxetable}

For consistency, we use the $\teff$ and [Fe/H] values only from SPC for TOI-4776 and TOI-5422 to set our priors for the global analysis discussed in the next section.

\section{Data Analysis} \label{sec:analysis}
\subsection{Global Fit with {\fontfamily{qcr}\selectfont EXOFASTv2}}
\label{sec:EXOFAST_fit}
We used {\fontfamily{qcr}\selectfont EXOFASTv2} for global analyses on TOI-4776 and TOI-5422 systems to derive parameters of the hosting stars and transiting BDs. The full description of {\fontfamily{qcr}\selectfont EXOFASTv2} is given in \cite{Eastman2019}. It uses the Monte Carlo Markov Chain (MCMC) method. For each MCMC fit, we use N = 36 (N = 2 × $n_{parameters}$) walkers, or chains, and run iteratively until the fit passes the default convergence criteria described in \cite{Eastman2019}.

Our inputs for {\fontfamily{qcr}\selectfont EXOFASTv2} and the obtained parameters are described here. Firstly, we input the stellar magnitudes referenced in \textbf{Table \ref{tab:SED_Mags}} and use the $\mathrm{T_{eff}}$ and [Fe/H] estimated by TRES as the priors on the SED model for each star. When deriving the stellar parameters, {\fontfamily{qcr}\selectfont EXOFASTv2} utilizes the MESA Isochrones and Stellar Tracks (MIST) model \citep{Paxton2015, Choi2016, Dotter2016}. We also input parallax measurements from Gaia DR3, after applying the zero-point corrections \citep{Lindegren2021}. These parallaxes are used with the SED model and an upper limit on V-band extinction ($A_V$; \cite{Schlafly2011}) to determine the stellar luminosity and radius. Then we use the determined stellar radius with the radius ratios obtained from transit photometry to constrain the radius of each transiting BD. These light curves also provide an inclination, which we combine with our input RV follow-up to constrain the mass and orbital properties of each transiting BD.

\begin{deluxetable}{@{}ccccc@{}}
\tablecaption{Magnitudes for TOI-4776 (TIC 196286578) and TOI-5422 (TIC 80611440). \label{tab:SED_Mags} }
\tabletypesize{\footnotesize}
\tablehead{
\colhead{} & \colhead{Description} & \colhead{TOI-4776} & \colhead{TOI-5422} & \colhead{Source}
}
\startdata
$\rm B_T$ & Tycho $\rm B_T$ & - & 12.634$\pm$0.317 & 1 \\
$\rm V_T$ & Tycho $\rm V_T$ & - & 11.599$\pm$0.168 & 1 \\
J2M & 2MASS J & 11.042$\pm$0.030 & 10.579$\pm$0.020 & 2, 3 \\
H2M & 2MASS H & 10.779$\pm$0.040 & 10.246$\pm$0.020 & 2, 3 \\
K2M & 2MASS $\rm K_s$ & 10.679$\pm$0.030 & 10.161$\pm$0.020 & 2, 3 \\
WISE1 & WISE 3.4$\mu$m & - & 10.135$\pm$0.030 & 4 \\
WISE2 & WISE 4.6$\mu$m & - & 10.173$\pm$0.030 & 4 \\
WISE3 & WISE 12$\mu$m & - & 10.209$\pm$0.087 & 4 \\
Gaia & Gaia G & 12.0280$\pm$0.0002 & 11.708$\pm$0.001 & 5 \\
GaiaBP & Gaia blue band & 12.3201$\pm$0.0011 & 12.080$\pm$0.003 & 5 \\
GaiaRP & Gaia red band & 11.5656$\pm$0.0011 & 11.184$\pm$0.001 & 5 \\
\enddata
\tablecomments{Tycho magnitudes are not used to model the SEDs and constrain $\teff$ for TOI-5422. The WISE1, 2, 3 magnitudes for TOI-4776 experience contamination from nearby stars, so they are not included in its SED fitting. \\
References. (1) \cite{TychoMag}; (2) \cite{UCAC4Mags}; (3) \cite{2MASSMags}; (4) \cite{ALLWISEMags}; (5) \cite{GaiaDR2Mags}.}
\end{deluxetable}

We set either uniform $\mathcal{U}$[a, b] or Gaussian $\mathcal{G}$[a, b] priors on our input parameters. We set Gaussian priors with mean $a$ and width $b$ on [Fe/H], $\teff$, and parallax, using the spectroscopic measurements of [Fe/H] and $\teff$ and parallax measurements from Gaia DR3 (see \textbf{Table \ref{tab:priors}}). For the TESS light curves of TOI-4776 only, we supplied an additional uniform prior for the dilution factor $A_D$. A detailed description of how priors are implemented in {\fontfamily{qcr}\selectfont EXOFASTv2} is given in \cite{Eastman2019}. 

\begin{deluxetable}{@{}cccc@{}}
\tablecaption{Priors used on the input parameters for {\fontfamily{qcr}\selectfont EXOFASTv2}.\label{tab:priors}}
\tabletypesize{\footnotesize}
\tablehead{ \colhead{} & \colhead{TOI-4776}  & \colhead{TOI-5422} & \colhead{Source} }
\startdata
$\mathrm{T_{eff}}$ & $\mathcal{G}${[}6008, 34{]}       & $\mathcal{G}${[}5736, 40{]} & TRES \\
{[}Fe/H{]} & $\mathcal{G}${[}-0.06, 0.06{]}     & $\mathcal{G}${[}-0.05, 0.05{]}   & TRES \\
Parallax   & $\mathcal{G}${[}2.69944, 0.0137{]} & $\mathcal{G}${[}2.87329, 0.0234{]} & Gaia DR3 \\
$A_V$      & $\mathcal{U}${[}0, 0.34255{]}      & $\mathcal{U}${[}0, 3.84214{]}   & 1 \\
$A_D$      & $\mathcal{U}${[}0.15, 0.25{]}      &    & \\
\enddata
\tablecomments{$\mathcal{U}$[a, b] and $\mathcal{G}$[a, b] represent uniform and Gaussian priors, respectively. Source 1: \cite{Schlafly2011}. $A_D$ is the dilution factor described in Section \ref{sec:TESS_lc4776}, which was only applied to the TESS light curves for TOI-4776.}
\end{deluxetable}

\subsubsection{Blended Light in the TOI-4776 System}
As mentioned in section \ref{sec:ground_lc4776}, there are nearby stars within the same TESS pixel as TOI-4776, contaminating the light curves. \textbf{Table \ref{tab:nearby_4776}} lists the nearby sources within 11'' of TOI-4776 from Gaia DR3 data. Since these sources are very close to our target, the magnitudes used for the star's SED fitting are contaminated. Given that the angular resolution for WISE1, 2, 3 are 6.1, 6.4, 6.5'', the WISE magnitudes for TOI-4776 given in \textbf{Table \ref{tab:SED_Mags}} did not pick the bright star TOI-592, but are contaminated by the 14.2 mag star with a separation of 4.04''. The additional light from the nearby 14.2 mag star, combined with the light from TOI-4776, makes the transit appear shallower than it actually is, which results in a smaller BD radius. Hence, for TOI-4776 only, we excluded the WISE1, 2, 3 magnitudes.

\begin{deluxetable*}{@{}cccccccc@{}}
\tablecaption{Nearby Sources (within 11'') to TOI-4776 from Gaia DR3 Data. The Gaia magnitude for TOI-4776 is 12.028. \label{tab:nearby_4776}}
\tabletypesize{\footnotesize}
\tablehead{ 
\colhead{TIC ID} & \colhead{RA} & \colhead{DEC} & \colhead{Distance} & \colhead{PM RA} &
\colhead{PM DEC} & \colhead{GAIA} & \colhead{Separation} \\ 
\colhead{} & \colhead{(J2015.5)} & \colhead{(J2015.5)} & \colhead{(pc)} & \colhead{(mas/yr)}
& \colhead{(mas/yr)} & \colhead{(Mag)} & \colhead{(arcsec)}
}
\startdata
832537467 & 08:22:12.8 & 08:22:12.8 & 1136.33 ± 32.055 & -6.37485 ± 0.0347 & 12.197 ± 0.03609 & 14.5889 & 4.04 \\
196286575 & 08:22:13.16 & -25:03:55.19 & 353.743 ± 15.732 & -37.1509 ± 0.1585 & 28.5776 ± 0.14324 & 17.3785 & 8.07 \\
TOI-592 & 08:22:13.72 & -25:04:10.09 & 358.441 ± 3.883 & 4.83617 ± 0.04321 & -7.165 ± 0.03996 & 10.5406 & 10.94 \\ 
\enddata
\end{deluxetable*}

\begin{figure*}[!h]
\centering
    \includegraphics[width=\textwidth]{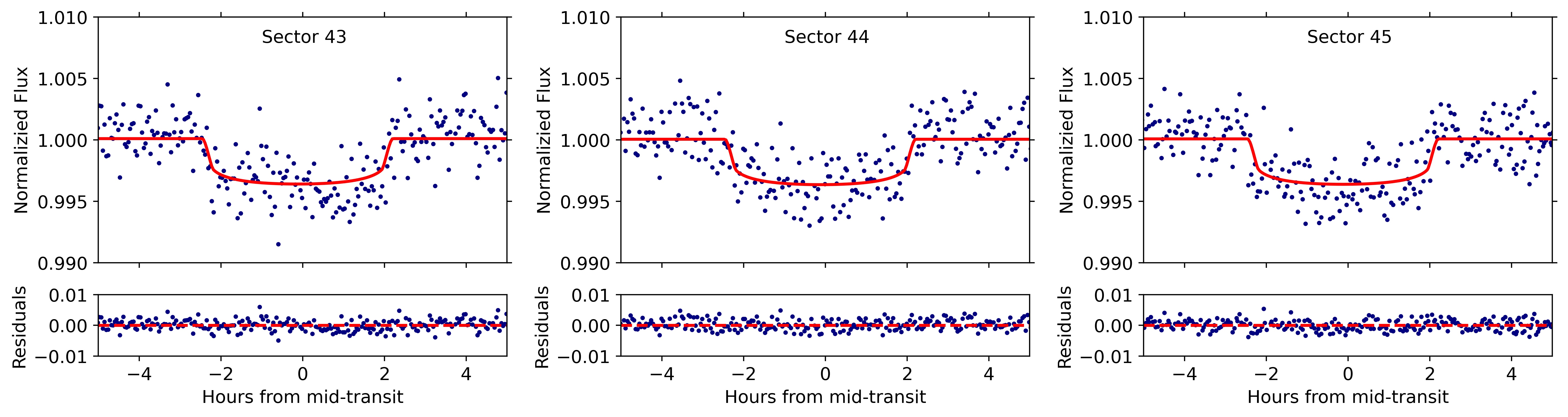}
    \includegraphics[width=0.667\textwidth]{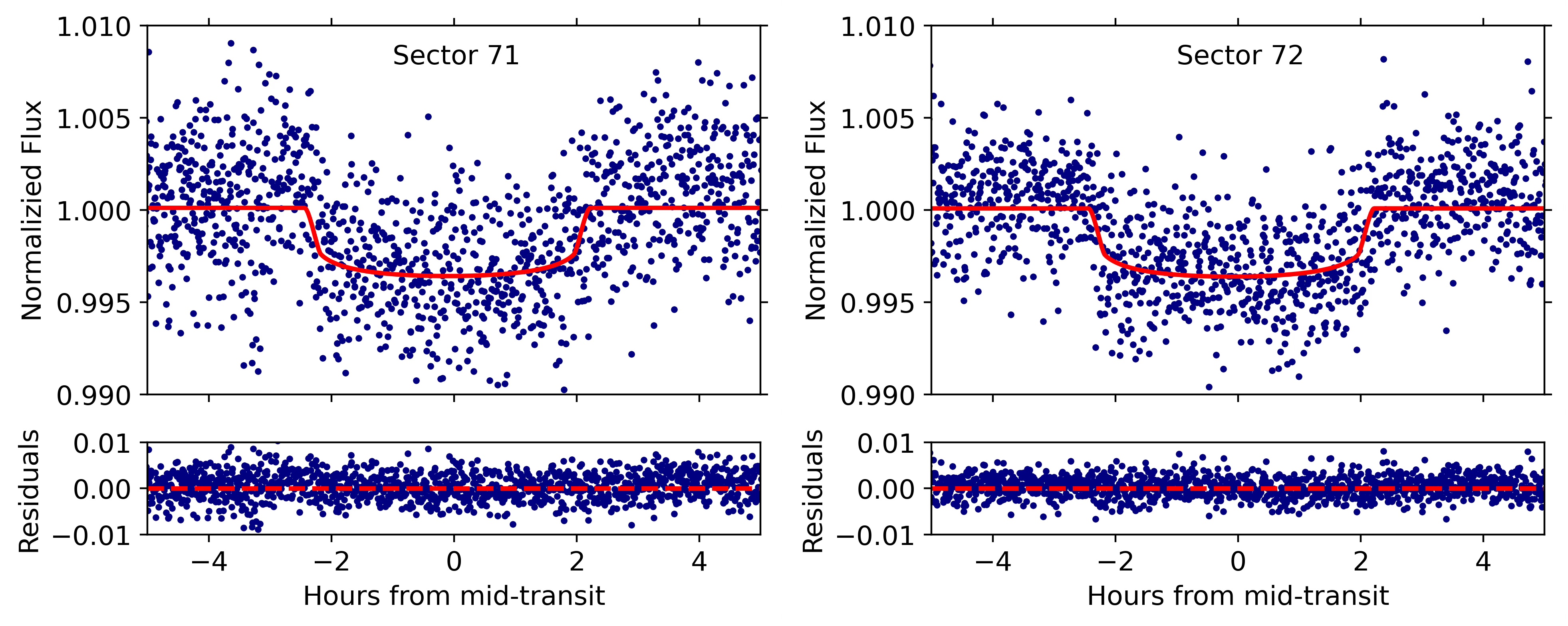}
    \caption{Phase-folded TESS light curves with residuals of each sector for TOI-5422. The red line is the fitted model from {\fontfamily{qcr}\selectfont EXOFASTv2}.}
\label{fig:fold_trans_all_5422}
\end{figure*}

\begin{figure} 
\centering
    \includegraphics[width=\columnwidth]{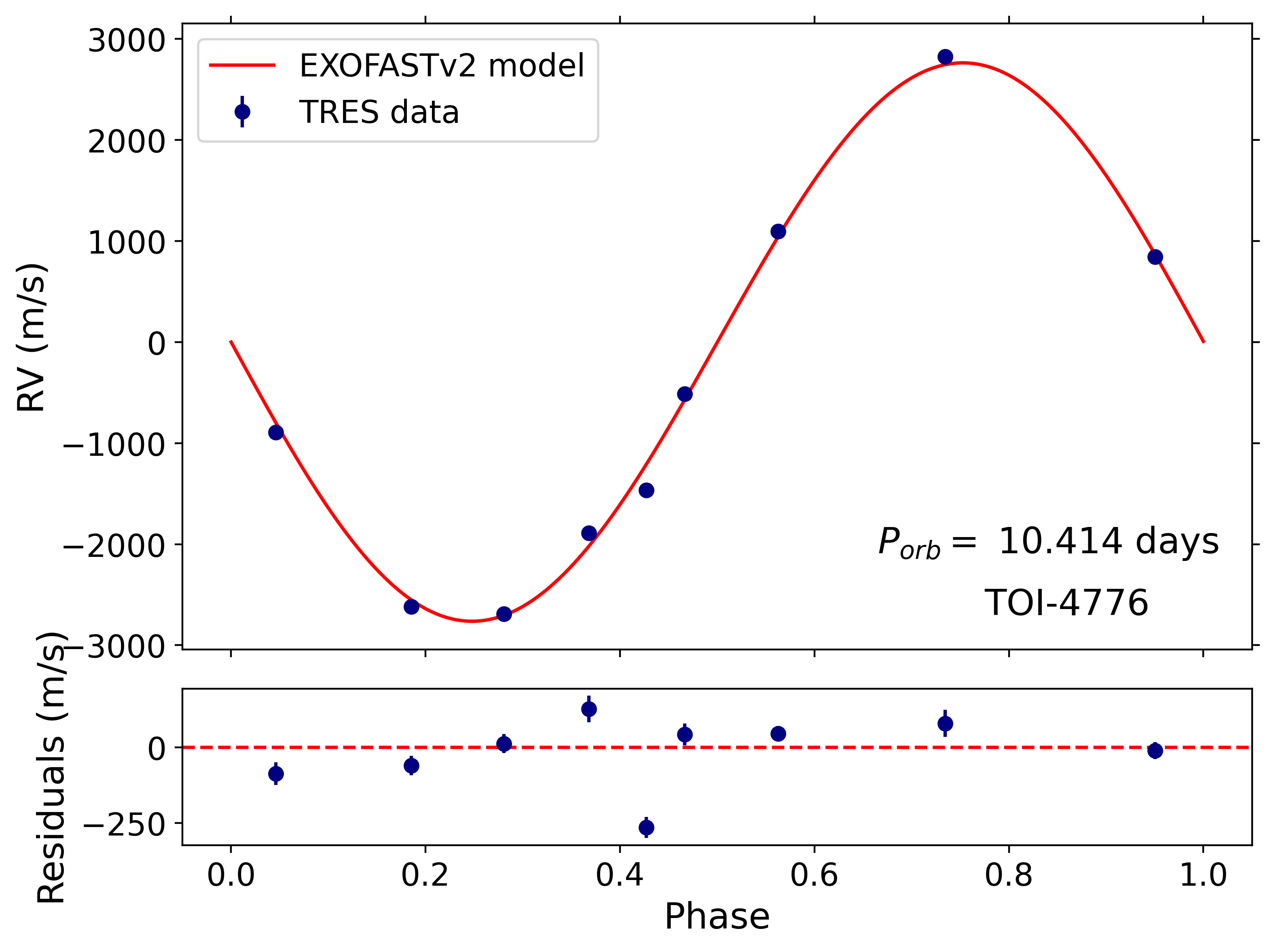}
    \hfill
    \includegraphics[width=\columnwidth]{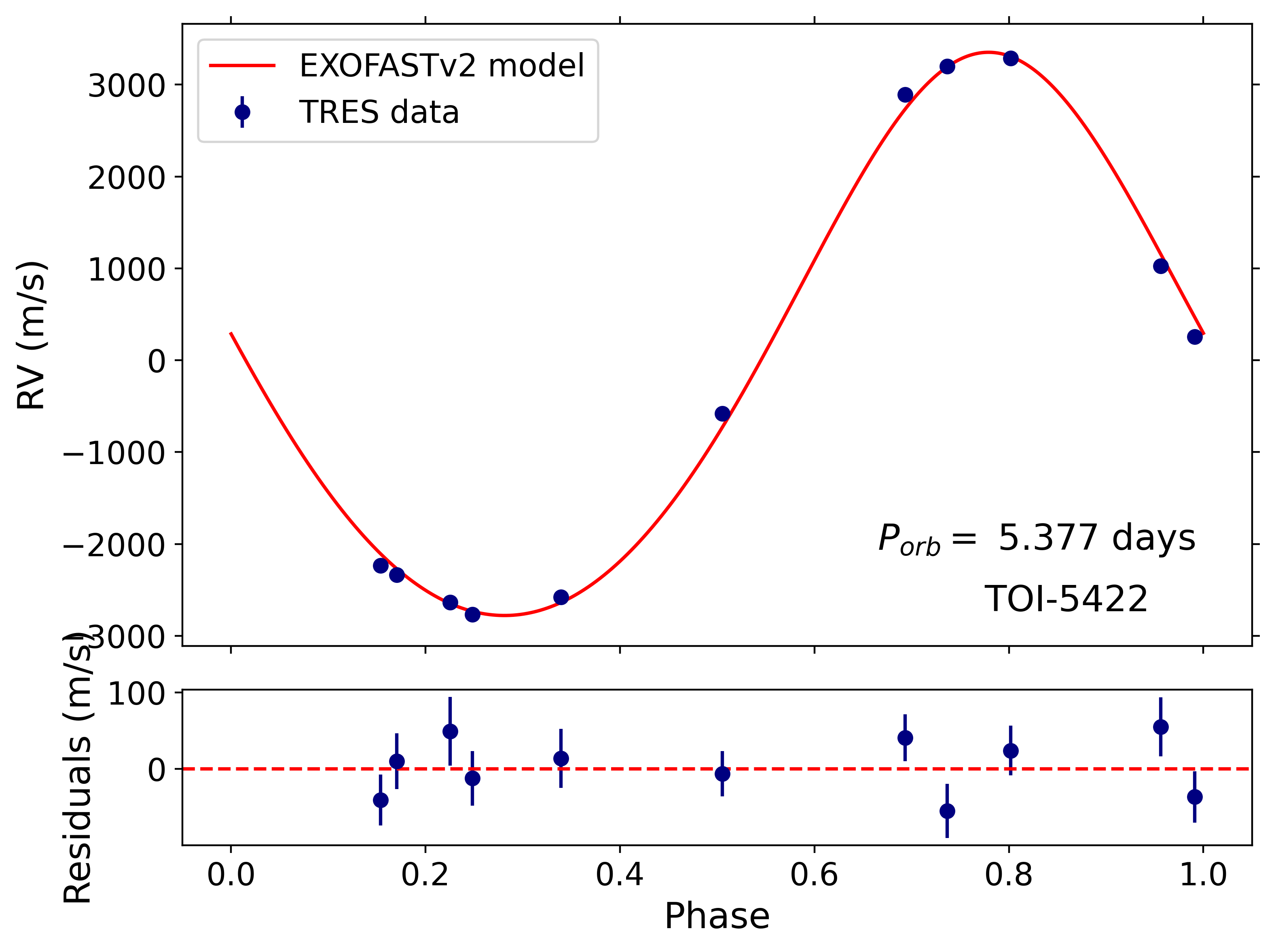}
    \centering
    \caption{Phase-folded radial velocity with residuals for TOI-4776 (top) and TOI-5422 (bottom). The red line is the fitted model from {\fontfamily{qcr}\selectfont EXOFASTv2}.}
\label{fig:fold_rv}
\end{figure}

\begin{figure}
\centering
    \includegraphics[trim={0.5cm 0cm 0.5cm 0.5cm}, clip, width=\columnwidth]{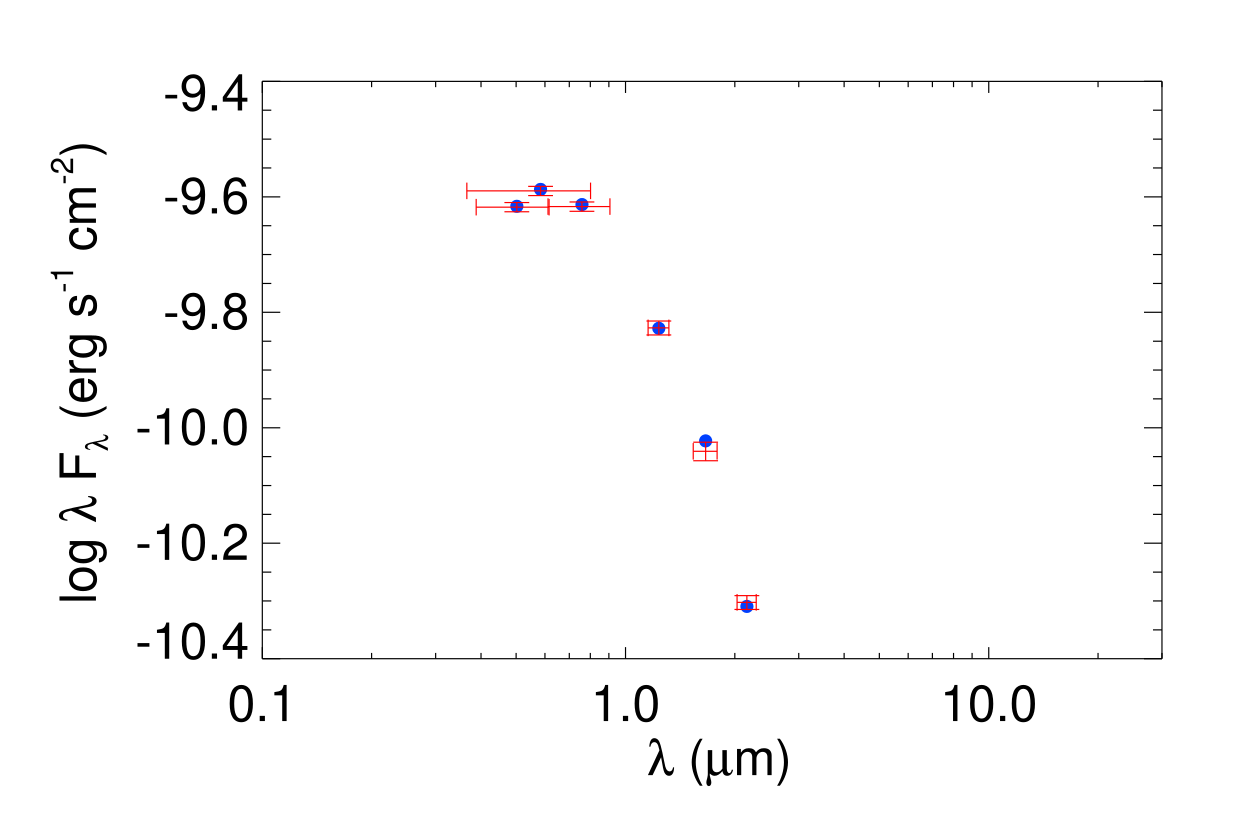}
    \includegraphics[trim={0.5cm 0cm 0.5cm 0.5cm}, clip, width=\columnwidth]{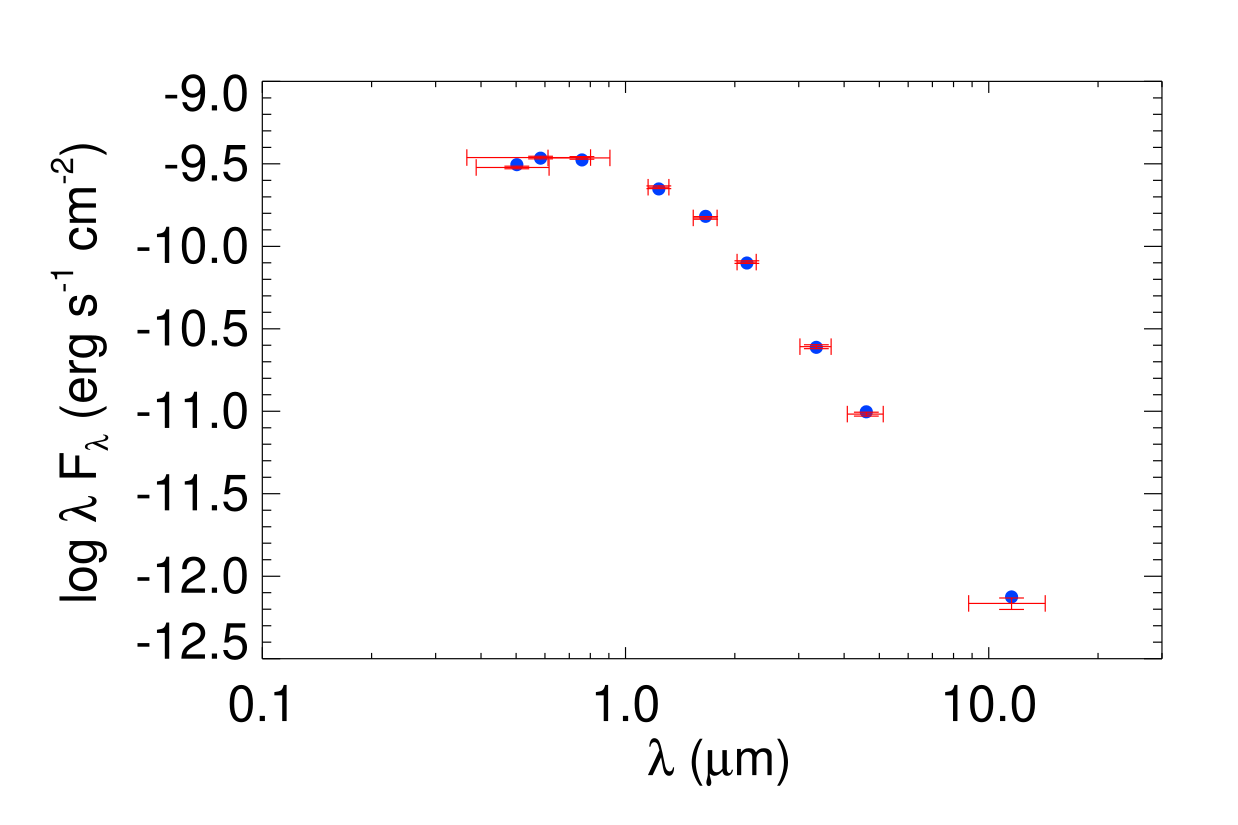}
    \centering
    \caption{SEDs computed using the MIST models built into {\fontfamily{qcr}\selectfont EXOFASTv2} for TOI-4776 (top) and TOI-5422 (bottom). The red symbols are the observed photometric measurements, where the horizontal bars represent the effective width of the bandpass. Blue symbols are the model values.}
\label{fig:starSED}
\end{figure}

The full set of derived parameters for each system is shown in \textbf{Tables \ref{tab:4776_fitresults}} and \textbf{Table \ref{tab:5422_fitresults}}. The transit model of each sector for TOI-5422 is shown in \textbf{Figure \ref{fig:fold_trans_all_5422}}, and the transit model for TOI-4776 is shown earlier in \textbf{Figure \ref{fig:4776_transit}}. The radial velocity data for both targets are shown in \textbf{Figure \ref{fig:fold_rv}}. The SED derived by {\fontfamily{qcr}\selectfont EXOFASTv2} for each star is shown in \textbf{Figure \ref{fig:starSED}}. 

\include{ResultTables.tex}

\subsection{Rotation Analysis for TOI-5422} \label{sec:5422_rotation}
\begin{figure}
    \centering
    \includegraphics[trim={0.6cm 0.3cm 1cm 1.5cm}, clip, width=\columnwidth]{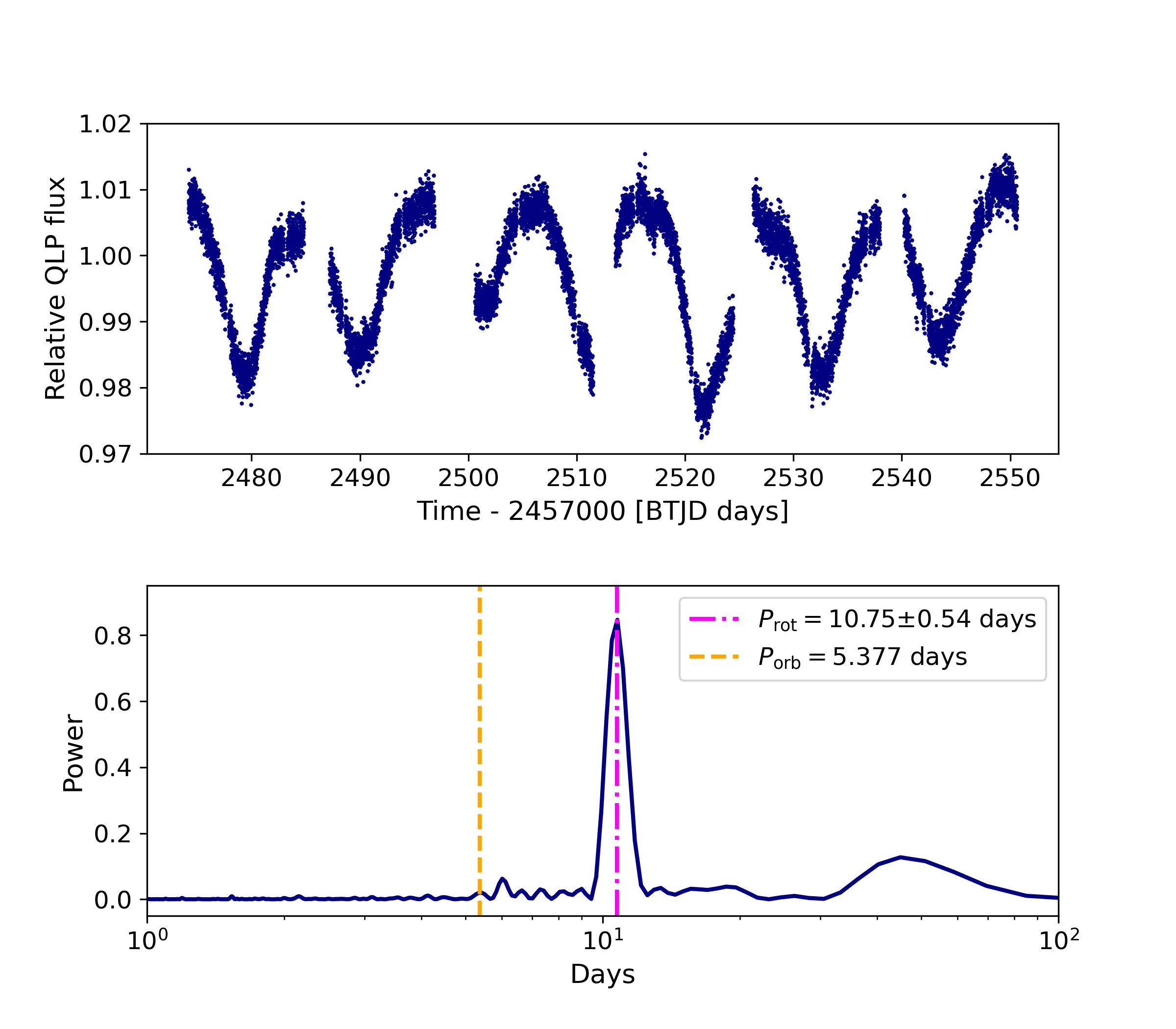}     
    \caption{\normalsize Top: The raw QLP light curves for TOI-5422 in sector 43, 44, 45, with all transits removed. There are clear photometric modulations caused by stellar rotation. Bottom: Lomb–Scargle periodogram of the data shown in the top panel. The magenta dash line is the resulted star's rotation period, and the orange dash line is the orbital period of the BD.}
\label{fig:5422_rotation}
\end{figure}

The raw TESS light curves of TOI-5422 reveal photometric modulations caused by stellar rotation, which motivates performing a rotational analysis. To characterize the periodicity of this modulation, we used the raw light curves of the 3 consecutive sectors, 43, 44, 45, and masked out the transits. Then, we performed a Lomb-Scargle periodogram analysis, shown in \textbf{Figure \ref{fig:5422_rotation}}. We see a peak frequency at 10.75$\pm$0.54 days. In order to make sure this obtained period is not an alias, we checked the phase-folded light curves and the inferred period from the projected stellar rotation velocity. Both checks seem to agree with this $\sim$10.75 day period. Since the BD's orbital period is shorter than the star's rotation period, the BD is likely spinning up the host star, making it rotate faster than subgiants with similar ages \citep{Jen2019}. Given the BD's orbital period of 5.377 days, roughly half of the star's rotation period, it is suggested that the rotation period of the star and the orbit period of the BD are very close to a 2:1 ratio. However, it is questionable to say that TOI-5422 and its companion BD are in a 2:1 spin-orbit resonance, since  typically only small rocky planets can maintain the permanent quadruple moment and spin–orbit resonance requires non-zero eccentricity \citep{Murray_Dermott_2000}. TOI-5422b is much more massive and has nearly zero eccentricity, so it is unlikely that TOI-5422b and its host star are in the case of 2:1 spin-orbit resonance.

\section{Discussion} \label{sec:discussion}

\subsection{Testing Young and Old Substellar Isochrones}
We can use transiting BDs with well-determined masses, radii, and ages to test a variety of substellar isochrones. Here we focused on the COND03 \citep{Baraffe2002}, ATMO 2020 \citep{ATMO2020}, and S24 \citep[Sonora Diamondback,][]{Diamondback} models. The COND03 models are appropriate for irradiated BDs and use the same input physics as \cite{ChabrierBaraffe1997} for main-sequence stars and scale them properly for low-mass stars and substellar objects down to $\lessapprox 1 \mj$. The ATMO 2020 models include the effects of clouds and have three cases, one using equilibrium chemistry (CEQ) and two using non-equilibrium chemistry due to strong or weak vertical mixing. And the S24 models consider the effect of both clouds and metallicity, including cloud parameter that characterizes thin to thick clouds ($f_{sed}=1-8$) and solar metallicity or two other abundance cases (Z = -0.5, 0.0, +0.5).

\begin{figure*}
\centering
    \includegraphics[width=0.8\textwidth]{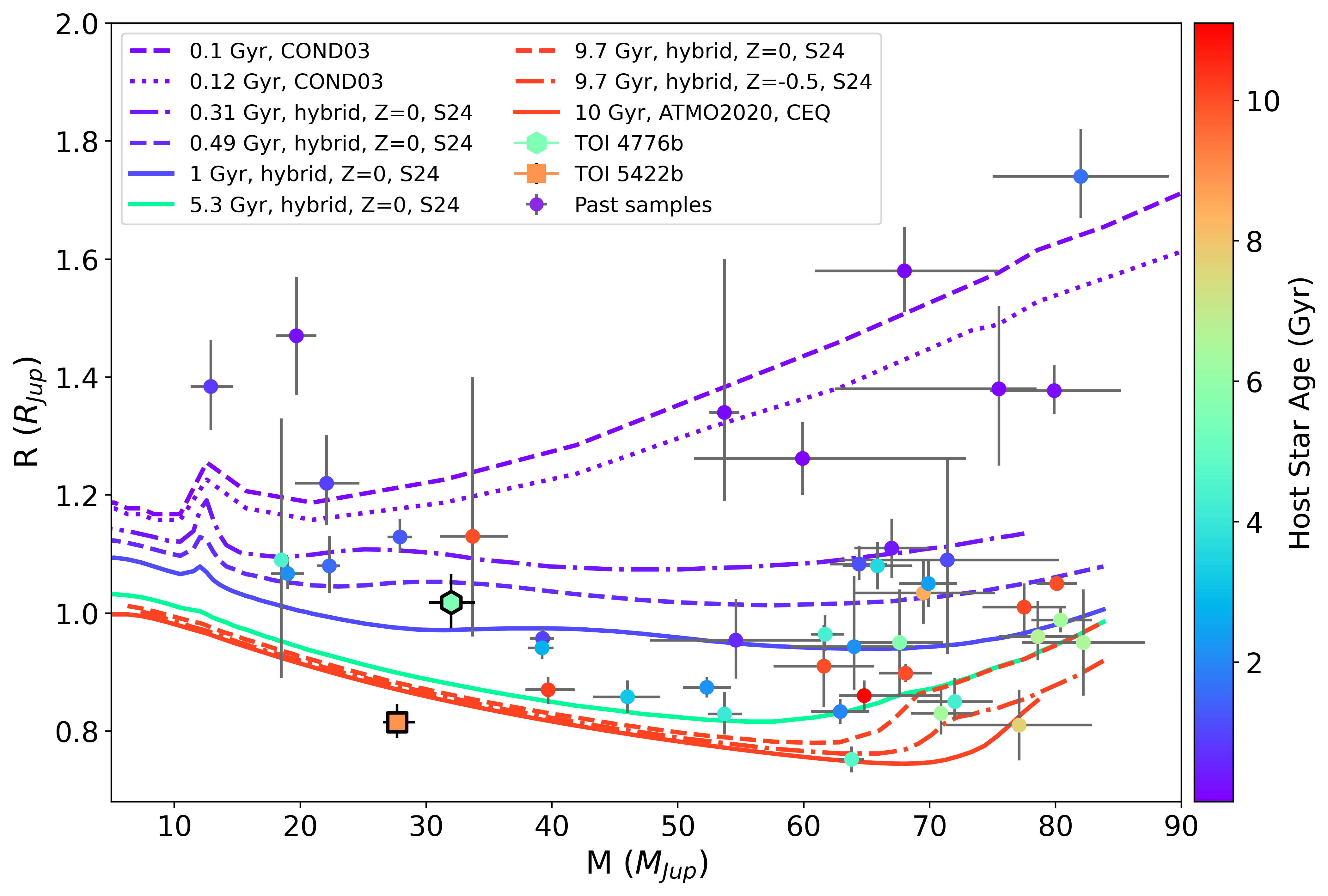}
    \caption{\normalsize The mass-radius diagram of recent published transiting BDs, color-coded by the host star's age, with different evolution isochrones overplotted. Our BD targets, TOI-4776b and TOI-5422b, are the large hexagon and square with black edges. Their locations, error bars, and color code are given by {\fontfamily{qcr}\selectfont EXOFASTv2} fit. The models are in the same colormap as the BD samples. }
\label{fig:evolution_models}
\end{figure*}

\textbf{Figure \ref{fig:evolution_models}} shows the mass-radius (M-R) relationship of a large fraction of known transiting BDs and our two targets with age color-coded by the host star's age, and it has different evolution models overplotted. The models are color-coded in the same way as the BD samples. The radius of BDs changes with its age, more rapidly at younger ages and then asymptotically approaching a constant value. Therefore, substellar isochrones becomes less effective predicting the age of BDs after a few Gyr because the rate at which BD radii contract significantly decelerates. This can be seen from \textbf{Figure \ref{fig:evolution_models}} where the $\sim$5 Gyr and $\sim$10 Gyr isochrones partially overlap with each other. The oldest substellar isochrones are traced better by a handful of transiting BDs than the younger ones, which might suggest that the oldest substellar isochrones predict the radii of transiting BDs more accurately as they approaching the asymptotic limit. Also, several BDs known to transit main-sequence stars seem to be inflated, e.g. KELT-1b \citep{KELT-1b}, GPX-1b \citep{GPX-1b}, and HATS-70b \citep{HATS-70b}.

We choose to plot the COND03 models for the 0.1 Gyr and 0.12 Gyr isochrones, as it covers a larger mass range than other substellar models for such young ages. The $\sim$0.3 Gyr, $\sim$0.5 Gyr, 1 Gyr, 5.3 Gyr, and 0.97 Gyr isochrones are all from S24, as it considers the effects of both clouds and metallicity. We chose the ``hybrid" case where clouds are included above 1300 K and cloud-free below 1300 K. The gentle hump at the middle masses are the result of different cooling rate above and below the L/T transition at 1300 K. Atmospheres cool slightly faster until the transition, and then stall when the clouds clear \citep{Diamondback}. Finally, the ATMO2020 model is used for the 10 Gyr isochrone, as it predicts the smallest radii for substellar objects at 10 Gyr, which grants us a wider range in radii to explore. 

\subsubsection{TOI-4776b and TOI-5422b in the Mass-Radius Diagram}
Our determinations for the mass, radius, and age for TOI-4776b are $32.0_{+1.9}^{-1.8}\mj$, $1.018^{+0.048}_{-0.043}\rj$, and $5.4^{+2.8}_{-2.2}$ Gyr, and TOI-5422b has $27.7^{+1.4}_{-1.1}\mj$, $0.815^{+0.031}_{-0.026}\rj$, and $8.2\pm2.4$ Gyr. As we can see in \textbf{Figure \ref{fig:evolution_models}}, TOI-4776b has an inflated radius with respect to the model prediction at the same mass and age. On the contrary, TOI-5422b is slightly below but relatively in agreement with the models, no matter which model and metallicity we choose.

The inflated radius of TOI-4776b is somewhat common among the younger transiting BD populations. This may be due to the stellar irradiation form its host star, causing the BD to puff up, or the cooling mechanisms for BDs are less effective over time than substellar models predict. In contrast, TOI-5422b is slightly below the 10 Gyr isochrone. This is interesting since no current transiting BDs lie below the 10 Gyr isochrone. Also, with the age of $8.2\pm2.4$ Gyr, TOI-5422b is among the oldest transiting BDs discovered. Note that for a fixed BD mass and age, the radius increases with higher-metallicity and higher-cloud-thickness atmospheres \citep{Burrows2011}. But in the case of our target, TOI-5422b, changing the metallicity of the model does not change the conclusion. \textbf{Figure \ref{fig:evolution_models}} shows that different metallicity values (Z = -0.5, 0.0) only makes noticeable difference beyond $\sim 60 \mj$ for the 9.7 Gyr isochrone. TOI-5422b is in the low-mass BD range, so even in the metal-poor scenario with Z = -0.5, it still lies slightly below the 10 Gyr isochrone. One possible explanation for its ``underluminous" radius is TOI-5422b being very metal-poor, so it is transparent to infrared light that would otherwise contribute to heating its atmosphere. However, we cannot get the BD's metallicity information from the data we have, and we cannot simply assume the BD has the same metallicity as the host star.

\subsubsection{Compare TOI-4776b with TOI-1406b}
We noticed an interesting fact that TOI-4776b has a `twin', TOI-1406b. TOI-1406b was discovered and characterized in \cite{Carmichael2020}. It has very similar host star and orbital properties with TOI-4776b, except it is $\sim$14$\mj$ more massive. TOI-1406b has an orbital period of $10.57415\pm0.00063$ days, a mass of $46.0\pm2.7 \mj$, and a radius of $0.86\pm0.03\rj$. Its host star has a mass of $1.18\pm0.09 M_{\odot}$, a radius of $1.35\pm0.03 R_{\odot}$, and an effective temperature of $\teff=6290\pm100$K. According to \cite{Carmichael2020}, the mass and radius of TOI-1406b are generally consistent with the model. Considering the effects of tidal heating on the BD's radius, if $a/R_{\star}>10$, the BD is `tidally detached' \citep{Rice2022}. For TOI-4776b, the effects of tidal heating can be ruled out, since it has $a/R_{\star}=16.93\pm0.66$; and this is the same case for TOI-1406b with $a/R_{\star}=16.11\pm0.58$. Therefore, it is reasonable to assume that TOI-1406b will become inflated, like the case for TOI-4776b, by just taking away a mass of $\sim$14$\mj$, i.e. reducing its surface gravity.

\subsection{Orbital Properties}
\subsubsection{Orbital Decay and Transit Time Variation for TOI-5422}
As shown in \textbf{Section \ref{sec:5422_rotation}}, the rotation period of the star in the TOI-5422 system is found to be 10.75$\pm$0.54 days, which is roughly twice the orbital period of the BD. When the orbital period of the BD is shorter than the rotational period of the star, the system is unstable. The BD deposits angular momentum onto the star, causing the star to spin up and the BD orbit to decay. To detect possible signatures of orbital decay, we used the TESS transit data from sector 43 (September 2021) and sector 72 (November 2023), which forms a baseline of 2 years. We individually fitted these two sectors with {\fontfamily{qcr}\selectfont EXOFASTv2} following the same procedure described in \textbf{Section \ref{sec:EXOFAST_fit}}. We obtained a period of $5.376767^{+0.000047}_{-0.000046}$ days for sector 43 and $5.376761\pm0.000046$ days for sector 72. Unfortunately, no obvious sign of orbital decay is detected for the TOI-5422 system. This failure of detection can be due to the relatively short 2-years baseline and the decay rate being too small. Transit data from future sectors is favored to help revealing possible signatures of orbital decay in this system.

\subsubsection{Rotational Inclination Angle for TOI-5422}
\label{sec:rot_incl}
\begin{figure}[!htb]
    \center
    \includegraphics[width=\columnwidth]{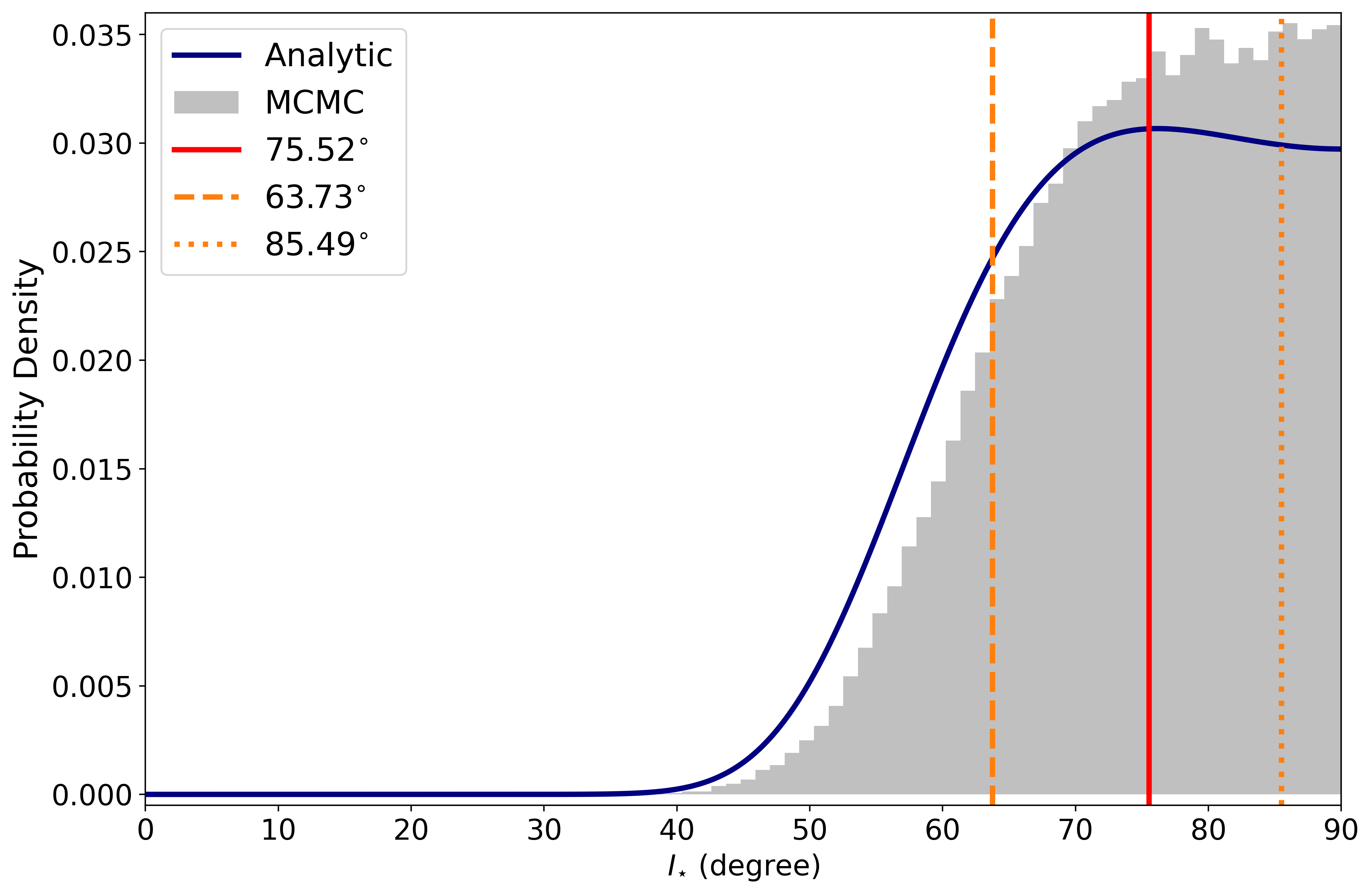}  
    \caption{\normalsize The analytic posterior distribution for the line-of-sight stellar inclination, $I_\star$, determined using the procedure in \cite{Bowler2023}, and the MCMC solution. The red vertical line corresponds to the median of the MCMC distribution, $I_{\star}=75.52^{\circ}$. Dashed orange lines are the 1-$\sigma$ credible interval (63.73$^{\circ}$, 85.49$^{\circ}$), which corresponds to the interval between the 16th and 84th percentiles of the MCMC solution. $I_\star=0^{\circ}$ corresponds to a star viewed pole-on, and $90^{\circ}$ is edge-on. }
\label{fig:steller_inclination_posterior}
\end{figure}

Since we detected photometric modulation in the light curve of TOI-5422, we can combine the information about the stellar rotation period within that modulation with the projected equatorial velocity of the star to determine the stellar inclination angle ${I_\star}$. ${I_\star}$ is the angle at which a star is inclined to the line of sight, which will help us to learn about the relative alignment between this angle and the orbital inclination angle, $i$, of a transiting or eclipsing object. The simplistic way of calculating ${I_\star}$ is
\begin{equation}
    I_\star = \sin^{-1} \left( \frac{v\sin{I_\star}}{V_{rot}} \right)
\label{equ:inclinationAngle}
\end{equation}
where $v\sin{I_\star}= 6.88 \pm 0.7$ $\rm km\, s^{-1}$ is the line-of-sight projection of the stellar rotation velocity derived from the TRES spectra by least-squares deconvolution(LSD) analysis \citep{Zhou2021}, and $V_{rot}=2 \pi R_{\star}/P_{rot}$ is the equatorial rotational velocity. Note that the $v\sin{I_\star}$ by LSD is used here instead of the one from SPC in \textbf{Section \ref{sec:RVdata}}, because the SPC radial velocity measurements are better for precise velocities and metallicities, while the LSD spectra are more appropriate for $v\sin{I_\star}$. We obtained $R_{\star}$ from our {\fontfamily{qcr}\selectfont EXOFASTv2} results and $P_{rot}$ from our Lomb–Scargle periodogram analysis in \textbf{Section \ref{sec:5422_rotation}}. However, \textbf{Equation \ref{equ:inclinationAngle}} should only be used with extra care since the priors on $v\sin{I_\star}$ and $V_{rot}$ are dependent on each other. Therefore, we must use statistical inference to calculate $I_\star$ to avoid biasing the $I_\star$ result and the lack of information on the measurement uncertainties. \cite{Masuda2020} and \cite{Bowler2023} outlined in details the procedure of leveraging the probability distribution of $\cos{I_\star}$ and $I_\star$ values. \textbf{Figure \ref{fig:steller_inclination_posterior}} shows our results of $I_\star$ using the formulation described in \cite{Bowler2023}. We obtained $I_{\star}=75.52^{+9.96}_{-11.79}$$^{\circ}$ for TOI-5422. We used the median of the MCMC distribution and calculated the 1-$\sigma$ uncertainties as the 16th and 84th percentiles of the distribution. Given that the orbital inclination of the system is $i=88.44^{+0.97}_{-0.86}$$^{\circ}$ by {\fontfamily{qcr}\selectfont EXOFASTv2} fit, it is likely that the BD's orbit is aligned with the stellar spin axis. Note that compared with the analytic solution, the MCMC distribution tends to go to a plateau at higher values closer to 90$^{\circ}$. This is expected for aligned system, while misaligned systems usually have a narrower distribution. The fact that the system is likely to be aligned and there is no outer companion is more consistent with the planet-like formation scenario \citep{Mordasini2008, Mordasini2012}.

\subsubsection{Circularization, Synchronization, and Spin-Orbit Alignment Timescales}
Tidal interactions between the host star and any type of companions will produce long-timescale evolution of the orbital elements and stellar rotation. There are three effects we consider here. Firstly, the orbit of the companion and the host star will start to circularize according to the circularization timescale. Then, the orbital period of the companion synchronizes with the host star's rotation within the synchronization timescale. Finally, the tidal forces tend to align the rotation axes of the companion and the host star characterized by the spin-orbit alignment timescale \footnote{It is possible for an eccentric BD to be aligned (zero obliquity), especially in cases where it formed in an aligned orbit.} \citep{Mazeh2008}. These timescales are influenced by the mass, radius, separation, and tidal quality factor $Q$ of both the host star and companion of a system. The tidal quality factor $Q$ describes about the tidally susceptibility of a star or planet. Following the formalism from \cite{Jackson2008}, the equations for the orbital circularization timescale for a close-in companion are
\begin{equation}
    \frac{1}{\tau_{circ,\star}}=\frac{171}{16} \sqrt{\frac{G}{M_{\star}}} \frac{R_{\star}^5 M_{BD}}{Q_{\star}} a^{-\frac{13}{2}}
\end{equation}
\begin{equation}
    \frac{1}{\tau_{circ,BD}}=\frac{63}{4}\frac{\sqrt{GM_{\star}^3}R_{BD}^5}{Q_{BD}M_{BD}} a^{-\frac{13}{2}}
\end{equation}
\begin{equation}
    \frac{1}{t_{circ}}=\frac{1}{\tau_{circ,\star}}+\frac{1}{\tau_{circ,BD}}
\label{equ:circ_time}
\end{equation}

\noindent where $t_{circ}$ is the circularization timescale, which is how long it takes for the orbital eccentricity of an object to decrease by an exponential factor. The semi-major axis is described by $a$, $M_{\star}$ is the stellar mass, $R_{\star}$ is the stellar radius, $M_{\rm BD}$ is the BD mass, $R_{\rm BD}$ is the BD radius, $Q_{\star}$ is the tidal quality factor for the star, and $Q_{BD}$ is the tidal quality factor for the BD. The assumptions of using \textbf{Equation \ref{equ:circ_time}} are listed in detail in \cite{Jackson2008}. In \textbf{Table \ref{tab:circ_time}}, we consider the circularization timescales for a range of values on the tidal quality factors, $Q_{\star}$ and $Q_{BD}$, for both systems. $Q_{BD}$ has a value as low as $10^{4.5}$ based on conclusions found by \cite{Beatty2018} on the constraint on $Q_{BD}$ for CWW 89Ab, a BD with 39$\mj$. The lowest value of $Q_{\star}$ is chosen to be $10^6$, since $Q_{\star}<10^6$ becomes nonphysical for solar-mass stars \citep{Ogilvie2004, Barnes2015}.

The tidal theory is discussed here to consider the current near circular orbit for both systems. For TOI-4776, the circularization timescale is much longer than the age of the system, regardless of which $Q_{\star}$ and $Q_{BD}$ combination is chosen. Therefore, it is likely that the BD was formed in a near circular orbit and underwent a low-eccentricity migration, unless tidal dissipation was extremely efficient. However, the circularization timescale for TOI-5422 is short enough for all combinations of $Q_{\star}$ and $Q_{BD}$ values to allow for tidal interactions to have circularized the orbit of the BD within the age of the system. Consequently, it is difficult to tell whether or not the BD was formed in a circular orbit. 

We calculated the synchronization timescale following the equation in \cite{Albrecht2012}:
\begin{equation}
    \frac{1}{\tau_{CE}} =\frac{1}{10\times10^9\ yr}q^2 \left( \frac{a/R_{\star}}{40}\right)^{-6}
\label{equ:syn_time}
\end{equation}
where $q$ is the companion-to-star mass ratio and $\tau_{CE}$ is the synchronization timescale for stars with convective envelopes. Using \textbf{Equation \ref{equ:syn_time}}, it requires $2.0_{-0.44}^{+0.41}$ Gyr for the TOI-5422 system to synchronize and $69.4_{-16.7}^{+21.5}$ Gyr for TOI-4776. The synchronization timescale for TOI-4776 is much longer than the age of the system, so we do not expect it to be synchronized yet. For TOI-5422, it is clear that the system is not synchronized, since the BD's orbit period is roughly half of the star's rotation period as mentioned in \textbf{Section \ref{sec:5422_rotation}}. Given that the synchronization timescale is shorter than the very old age of TOI-5422, it is interesting to see that the BD has near circular orbit but still has not synchronized its orbit with the star’s rotation. One possible explanation can be that the BD formed in a very eccentric orbit and migrated inward through high-eccentricity migration. The BD circularized its orbit within the age of the system, and just finished circularizing and we now happens to see it synchronizing.

\begin{deluxetable}{cccc}
\tablecaption{Circularization timescales for different values of $Q_\star$ and $Q_{BD}$. \label{tab:circ_time}}
\tabletypesize{\small}
\tablehead{
\colhead{Object Name and Age} & \colhead{$Q_\star$} & \colhead{$Q_{BD}$} & \colhead{$t_{circ}$ (Gyr)}
}
\startdata
\multirow{4}{*}{\begin{tabular}[c]{@{}l@{}}TOI-4776\\ $5.4^{+2.8}_{-2.2}$ Gyr \end{tabular}}
 & $10^7$  & $10^6$     & $192.3_{-37.4}^{+44.4}$ \\
 & $10^7$  & $10^{4.5}$ & $57.6_{-12.0}^{+14.4}$ \\
 & $10^6$  & $10^{4.5}$ & $16.5_{-3.2}^{+3.8}$ \\ \hline
\multirow{4}{*}{\begin{tabular}[c]{@{}l@{}}TOI-5422 \\ $8.2\pm2.4$ Gyr \end{tabular}}
 & $10^7$  & $10^6$     & $5.1_{-0.9}^{+0.8}$ \\
 & $10^7$  & $10^{4.5}$ & $3.6_{-0.7}^{+0.5}$ \\
 & $10^6$  & $10^{4.5}$ & $0.5_{-0.1}^{+0.1}$ \\
\enddata
\end{deluxetable}

\subsection{Independent Determination of Stellar Parameters for TOI-4776}
\begin{figure}[!htb]
    \center
    \includegraphics[width=\columnwidth]{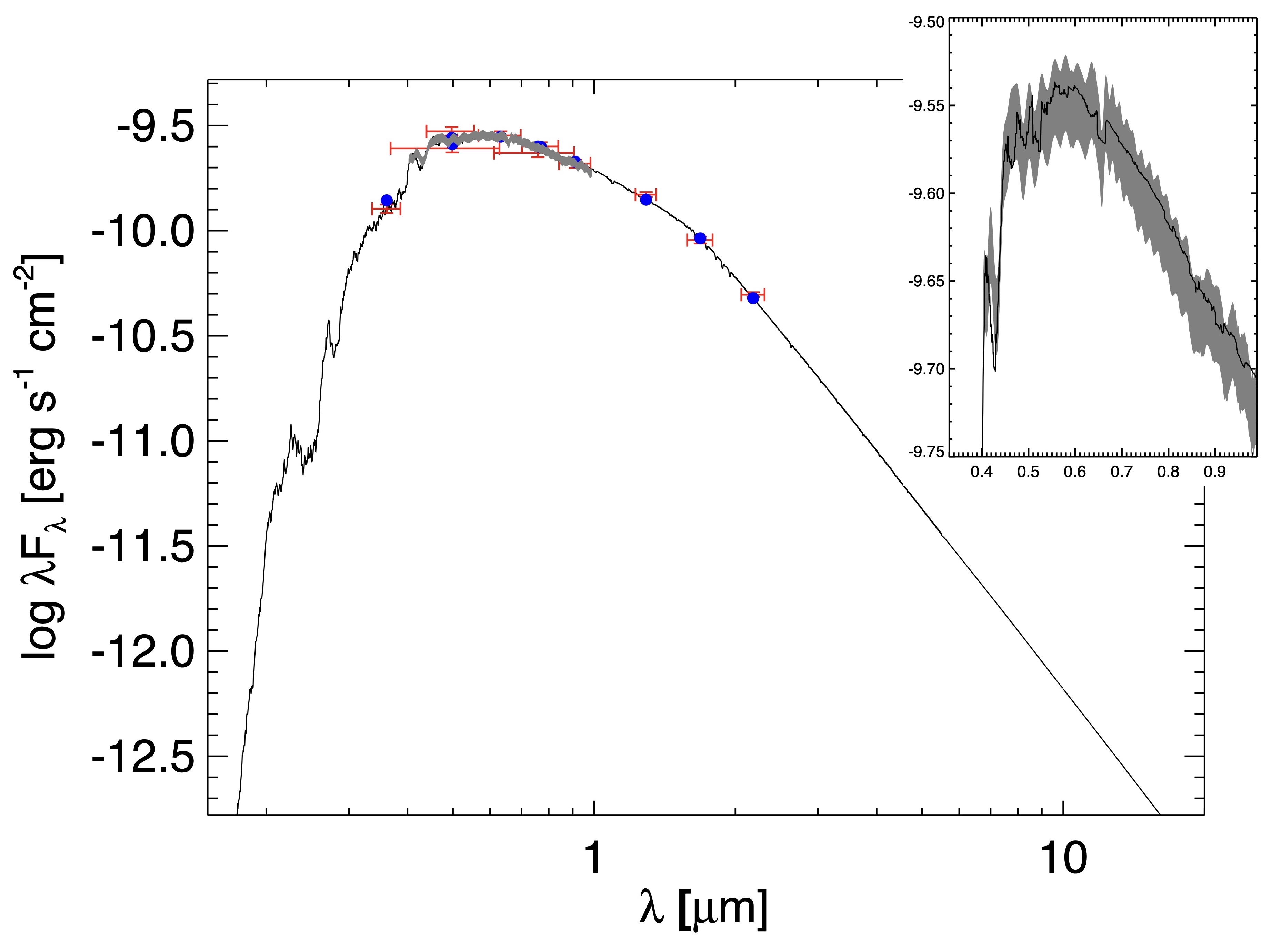}  
    \caption{Spectral energy distribution of TOI-4776. Red symbols represent the observed photometric measurements, where the horizontal bars represent the effective width of the passband. Blue symbols are the model fluxes from the best-fit PHOENIX atmosphere model (black). The inset shows the {\it Gaia\/} spectrophotometry overlaid as a grey swathe.}
\label{fig:TOI4776_sed_sep}
\end{figure}

As an independent determination of the basic stellar parameters, we performed an analysis of the broadband spectral energy distribution (SED) of the star together with the {\it Gaia\/} DR3 parallax \citep[with no systematic offset applied; see, e.g.,][]{StassunTorres:2021}, in order to determine an empirical measurement of the stellar radius, following the procedures described in \citet{Stassun:2016,Stassun:2017,Stassun:2018}. We pulled the $JHK_S$ magnitudes from {\it 2MASS}, the $G_{\rm BP} G_{\rm RP}$ magnitudes from {\it Gaia}, and the $ugriz$ magnitudes from Sky Mapper. We avoided the W1--W3 magnitudes from {\it WISE}, due to the contamination of that photometry by a nearby moderately bright source. We also utilized the absolute flux-calibrated {\it Gaia\/} spectrophotometry. Together, the available photometry spans the full stellar SED over the wavelength range 0.4--2~$\mu$m (see Figure~\ref{fig:TOI4776_sed_sep}).  
 
We performed a fit using PHOENIX stellar atmosphere models \citep{Husser:2013}, with the free parameters being the effective temperature ($T_{\rm eff}$), surface gravity ($\log g$), metallicity ([Fe/H]), and the extinction $A_V$, the latter of which was limited to maximum line-of-sight value from the Galactic dust maps of \citet{Schlegel:1998}. The resulting fit (Figure \ref{fig:TOI4776_sed_sep}) has a a reduced $\chi^2$ of 1.7, with a best-fit $A_V = 0.09 \pm 0.02$, $T_{\rm eff} = 6130 \pm 100$~K, $\log g = 4.3 \pm 0.3$, and [Fe/H] = $-0.3 \pm 0.3$. Integrating the (unreddened) model SED gives the bolometric flux at Earth, $F_{\rm bol} = 4.244 \pm 0.099 \times 10^{-10}$ erg~s$^{-1}$~cm$^{-2}$. Taking the $F_{\rm bol}$ together with the {\it Gaia\/} parallax directly gives the bolometric luminosity, $L_{\rm bol} = 1.858 \pm 0.044$~L$_\odot$. The stellar radius follows from the Stefan-Boltzmann relation, giving $R_\star = 1.209 \pm 0.042$~R$_\odot$. In addition, we can estimate the stellar mass from the empirical relations of \citet{Torres:2010}, giving $M_\star = 1.09 \pm 0.07$~M$_\odot$. 

Comparing these independently determined stellar parameters with our {\fontfamily{qcr}\selectfont EXOFASTv2} global fit results, they agree to each other within 1$\sigma$. Therefore, our decision to remove the 3 WISE magnitudes in TOI-4776's SED during the {\fontfamily{qcr}\selectfont EXOFASTv2} fit did not hurt our analysis, which is expected as the WISE magnitudes only occupy a small fraction of the wavelength range of the SED. Thus, this independent SED fitting serves as a confirmation of our previous analysis.

\section{Summary} \label{sec:summary}
TOI-4776b and TOI-5422b are two newly discovered low-mass BDs by TESS that transit their host stars in nearly circular orbits. We used TESS and ground-based light curves, ground-based RV follow-up, and parallax measurements from Gaia DR3 to characterize the BDs. Both BDs have masses that lie in the ``brown dwarf desert", but they have different ages. The younger TOI-4776 system is found to be $5.4^{+2.8}_{-2.2}$ Gyr old, with a late-F host star. The BD has a 10.4138$\pm$0.000014 day period, a mass of $32.0^{+1.9}_{-1.8}\mj$, and an inflated radius of $1.018^{+0.048}_{-0.043}\rj$. The older TOI-5422 system has a age of $8.2\pm2.4$ Gyr, making TOI-5422b among the oldest transiting BDs discovered. The BD has $27.7^{+1.4}_{-1.1}\mj$ with a smaller radius of $0.815^{+0.031}_{-0.026}\rj$, in a 5.377219$\pm$0.000010 day orbit. Its host star, TOI-5422, is a subgiant that exhibits periodic variations caused by rotation in its light curves. The rotation analysis show a stellar rotation period of 10.75$\pm$0.54 day, so the ratio between the stellar rotation period and the BD's orbital period is close to 2:1 and the BD is likely spinning up the host star. The stellar inclination angle is found to be $I_{\star}=75.52^{+9.96}_{-11.79}$$^{\circ}$, and the BD's orbit is expected to be aligned with stellar spin axis. Considering the different circularization timescales for both systems, TOI-4776b is expected to form in a near circular orbit, while it is difficult to tell whether or not TOI-5422b was formed in a circular orbit. TOI-4776 is not expected to be synchronized at this stage, and it is intriguing that TOI-5422, despite its old age, is also not synchronized.

\section{Future} \label{sec:future}
\subsection{The RV Jitter for TOI-4776}
The RV jitter term for TOI-4776 (late-F) was found to be $167^{+94}_{-51}$m/s by {\fontfamily{qcr}\selectfont EXOFASTv2}, while TOI-5422 (subgiant) has only $42^{+30}_{-22}$m/s. Here, we discuss the possible scenario that may cause the large RV jitters in TOI-4776. Usually, a large jitter term will suggest the existence of another planet or caused by asteroseismology. However, the jitter term for TOI-4776 is larger than the possible values produced by another planet or asteroseismology. If TOI-4776 is above Kraft break, it might explain the high jitter. It is also possible that TOI-4776 is just more active than the sub-giant. We only used 9 RV data points, so the number of RV points we have is not enough for another periodogram analysis to find if there is another planet. Therefore, we encourage future work investigating this problem.

\subsection{The Rossiter–McLaughlin Effect}
Both the TOI-5422 and TOI-4776 systems are amenable to Rossiter–McLaughlin (RM) observation, which is the most effective way to measure the sky-projected obliquity of the system. With the sky-projected obliquity, if the inclination of the stellar spin axis is known, we can calculate the true stellar obliquity. As we mentioned in \textbf{Section \ref{sec:rot_incl}}, TOI-5422b is likely to be aligned based on the calculation of the stellar inclination angle, but we do not have information about the obliquity of the system. For TOI-4776, we are not able to preform the Lomb–Scargle periodogram analysis to get the star's rotation period, so we do not know whether the BD is aligned with the host star's spin axis. Future work to obtain RM effect measurements would enable us to get the full 3D orbital configuration of the TOI-5422 system and only as much as the projected orbital obliquity, $\lambda$, for TOI-4776 as it lacks a measure stellar spin-axis inclination.

\section{Acknowledgements}
Funding for the TESS mission is provided by NASA's Science Mission Directorate. This paper includes data collected by the TESS mission that are publicly available from the Mikulski Archive for Space Telescopes (MAST). We acknowledge the use of public TESS data from pipelines at the TESS Science Office and at the TESS Science Processing Operations Center. Resources supporting this work were provided by the NASA High-End Computing (HEC) Program through the NASA Advanced Supercomputing (NAS) Division at Ames Research Center for the production of the SPOC data products. This research has made use of the Exoplanet Follow-up Observation Program website, which is operated by the California Institute of Technology, under contract with the National Aeronautics and Space Administration under the Exoplanet Exploration Program. 
The work of HPO has been carried out within the framework of the NCCR PlanetS supported by the Swiss National Science Foundation under grants 51NF40\_182901 and 51NF40\_205606.
This article is based on observations made with the MuSCAT2 instrument, developed by ABC, at Telescopio Carlos S\'{a}nchez operated on the island of Tenerife by the IAC in the Spanish Observatorio del Teide. This work is partly financed by the Spanish Ministry of Economics and Competitiveness through grants PGC2018-098153-B-C31. This work is partly supported by JSPS KAKENHI Grant Numbers
JP24H00017, JP24K00689, and JSPS Bilateral Program Number
JPJSBP120249910.

\bibliography{citation}{}
\bibliographystyle{aasjournal}

\end{document}

%% file: ResultTables.tex
\providecommand{\bjdtdb}{\ensuremath{\rm {BJD_{TDB}}}}
\providecommand{\feh}{\ensuremath{\left[{\rm Fe}/{\rm H}\right]}}
\providecommand{\teff}{\ensuremath{T_{\rm eff}}}
\providecommand{\teq}{\ensuremath{T_{\rm eq}}}
\providecommand{\ecosw}{\ensuremath{e\cos{\omega_*}}}
\providecommand{\esinw}{\ensuremath{e\sin{\omega_*}}}
\providecommand{\msun}{\ensuremath{\,M_\Sun}}
\providecommand{\rsun}{\ensuremath{\,R_\Sun}}
\providecommand{\lsun}{\ensuremath{\,L_\Sun}}
\providecommand{\mj}{\ensuremath{\,M_{\rm J}}}
\providecommand{\rj}{\ensuremath{\,R_{\rm J}}}
\providecommand{\me}{\ensuremath{\,M_{\rm E}}}
\providecommand{\re}{\ensuremath{\,R_{\rm E}}}
\providecommand{\fave}{\langle F \rangle}
\providecommand{\fluxcgs}{10$^9$ erg s$^{-1}$ cm$^{-2}$}
\startlongtable
\begin{deluxetable*}{lcc}
\tablecaption{Median values and 68\% confidence interval for TOI4776. \label{tab:4776_fitresults}}
\tablehead{\colhead{~~~Parameter} & \colhead{Units} & \multicolumn{1}{c}{Values}}
\startdata
\smallskip\\\multicolumn{2}{l}{Stellar Parameters:}&\smallskip\\
~~~~$M_*$\dotfill &Mass (\msun)\dotfill &$1.063^{+0.070}_{-0.068}$\\
~~~~$R_*$\dotfill &Radius (\rsun)\dotfill &$1.220^{+0.046}_{-0.040}$\\
~~~~$R_{*,SED}$\dotfill &Radius$^{1}$ (\rsun)\dotfill &$1.204^{+0.023}_{-0.022}$\\
~~~~$L_*$\dotfill &Luminosity (\lsun)\dotfill &$1.75^{+0.14}_{-0.11}$\\
~~~~$F_{Bol}$\dotfill &Bolometric Flux (cgs)\dotfill &$0.000000000408^{+0.000000000031}_{-0.000000000025}$\\
~~~~$\rho_*$\dotfill &Density (cgs)\dotfill &$0.823^{+0.10}_{-0.093}$\\
~~~~$\log{g}$\dotfill &Surface gravity (cgs)\dotfill &$4.291^{+0.039}_{-0.040}$\\
~~~~$T_{\rm eff}$\dotfill &Effective Temperature (K)\dotfill &$6011\pm33$\\
~~~~$T_{\rm eff,SED}$\dotfill &Effective Temperature$^{1}$ (K)\dotfill &$6050^{+130}_{-110}$\\
~~~~$[{\rm Fe/H}]$\dotfill &Metallicity (dex)\dotfill &$-0.050^{+0.051}_{-0.048}$\\
~~~~$[{\rm Fe/H}]_{0}$\dotfill &Initial Metallicity$^{2}$ \dotfill &$0.012^{+0.049}_{-0.045}$\\
~~~~$Age$\dotfill &Age (Gyr)\dotfill &$5.4^{+2.8}_{-2.2}$\\
~~~~$A_V$\dotfill &V-band extinction (mag)\dotfill &$0.123^{+0.087}_{-0.075}$\\
~~~~$\sigma_{SED}$\dotfill &SED photometry error scaling \dotfill &$0.88^{+0.57}_{-0.29}$\\
~~~~$\varpi$\dotfill &Parallax (mas)\dotfill &$2.700\pm0.030$\\
~~~~$d$\dotfill &Distance (pc)\dotfill &$370.4^{+4.2}_{-4.1}$\\
\smallskip\\\multicolumn{2}{l}{Planetary Parameters:}&b\smallskip\\
~~~~$P$\dotfill &Period (days)\dotfill &$10.413815\pm0.000014$\\
~~~~$R_P$\dotfill &Radius (\rj)\dotfill &$1.018^{+0.048}_{-0.043}$\\
~~~~$M_P$\dotfill &Mass (\mj)\dotfill &$32.0^{+1.9}_{-1.8}$\\
~~~~$T_C$\dotfill &Time of conjunction$^{4}$ (\bjdtdb)\dotfill &$2460016.5936\pm0.0011$\\
~~~~$T_T$\dotfill &Time of minimum projected separation$^{5}$ (\bjdtdb)\dotfill &$2460016.5936\pm0.0011$\\
~~~~$T_0$\dotfill &Optimal conjunction Time$^{6}$ (\bjdtdb)\dotfill &$2459527.14432^{+0.00091}_{-0.00092}$\\
~~~~$a$\dotfill &Semi-major axis (AU)\dotfill &$0.0962^{+0.0020}_{-0.0021}$\\
~~~~$i$\dotfill &Inclination (Degrees)\dotfill &$87.74\pm0.21$\\
~~~~$e$\dotfill &Eccentricity \dotfill &$0.014^{+0.023}_{-0.010}$\\
~~~~$\omega_*$\dotfill &Argument of Periastron (Degrees)\dotfill &$80^{+110}_{-120}$\\
~~~~$e\cos{\omega_*}$\dotfill & \dotfill &$0.0003^{+0.010}_{-0.0087}$\\
~~~~$e\sin{\omega_*}$\dotfill & \dotfill &$0.002^{+0.025}_{-0.012}$\\
~~~~$M_P\sin i$\dotfill &Minimum mass (\mj)\dotfill &$32.0^{+1.9}_{-1.8}$\\
~~~~$M_P/M_*$\dotfill &Mass ratio \dotfill &$0.0288\pm0.0012$\\
~~~~$T_{eq}$\dotfill &Equilibrium temperature$^{7}$ (K)\dotfill &$1032^{+21}_{-19}$\\
~~~~$\fave$\dotfill &Incident Flux (\fluxcgs)\dotfill &$0.258^{+0.021}_{-0.019}$\\
~~~~$\tau_{\rm circ}$\dotfill &Tidal circularization timescale (Gyr)\dotfill &$2490^{+670}_{-550}$\\
~~~~$K$\dotfill &RV semi-amplitude (m/s)\dotfill &$2806^{+98}_{-97}$\\
~~~~$R_P/R_*$\dotfill &Radius of planet in stellar radii \dotfill &$0.0857\pm0.0016$\\
~~~~$a/R_*$\dotfill &Semi-major axis in stellar radii \dotfill &$16.93\pm0.66$\\
~~~~$\delta$\dotfill &$\left(R_P/R_*\right)^2$ \dotfill &$0.00735\pm0.00028$\\
~~~~$\tau$\dotfill &Ingress/egress transit duration (days)\dotfill &$0.0225^{+0.0025}_{-0.0022}$\\
~~~~$T_{14}$\dotfill &Total transit duration (days)\dotfill &$0.1674^{+0.0027}_{-0.0025}$\\
~~~~$b$\dotfill &Transit Impact parameter \dotfill &$0.664^{+0.038}_{-0.044}$\\
~~~~$b_S$\dotfill &Eclipse impact parameter \dotfill &$0.670^{+0.038}_{-0.040}$\\
~~~~$\rho_P$\dotfill &Density (cgs)\dotfill &$37.5^{+5.8}_{-5.2}$\\
~~~~$logg_P$\dotfill &Surface gravity \dotfill &$4.883^{+0.046}_{-0.047}$\\
~~~~$\Theta$\dotfill &Safronov Number \dotfill &$5.68^{+0.34}_{-0.33}$\\
~~~~$T_P$\dotfill &Time of Periastron (\bjdtdb)\dotfill &$2460016.4^{+3.2}_{-3.3}$\\
~~~~$T_S$\dotfill &Time of eclipse (\bjdtdb)\dotfill &$2460011.389^{+0.068}_{-0.058}$\\
~~~~$T_A$\dotfill &Time of Ascending Node (\bjdtdb)\dotfill &$2460013.999^{+0.097}_{-0.052}$\\
~~~~$T_D$\dotfill &Time of Descending Node (\bjdtdb)\dotfill &$2460019.190^{+0.049}_{-0.081}$\\
~~~~$d/R_*$\dotfill &Separation at mid transit \dotfill &$16.82^{+0.78}_{-0.80}$\\
~~~~$P_T$\dotfill &A priori non-grazing transit prob \dotfill &$0.0543^{+0.0027}_{-0.0024}$\\
~~~~$P_{T,G}$\dotfill &A priori transit prob \dotfill &$0.0645^{+0.0033}_{-0.0029}$\\
~~~~$P_S$\dotfill &A priori non-grazing eclipse prob \dotfill &$0.0537^{+0.0024}_{-0.0023}$\\
~~~~$P_{S,G}$\dotfill &A priori eclipse prob \dotfill &$0.0638^{+0.0029}_{-0.0028}$\\
\smallskip\\\multicolumn{2}{l}{Wavelength Parameters:}& z'\\
~~~~$u_{1}$\dotfill &linear limb-darkening coeff \dotfill &$0.228^{+0.048}_{-0.049}$\\
~~~~$u_{2}$\dotfill &quadratic limb-darkening coeff \dotfill &$0.299\pm0.049$\\
\multicolumn{2}{l}{}&TESS\\
~~~~$u_{1}$\dotfill &linear limb-darkening coeff \dotfill &$0.261\pm0.025$\\
~~~~$u_{2}$\dotfill &quadratic limb-darkening coeff \dotfill & $0.292\pm0.025$\\
~~~~$A_D$\dotfill &Dilution from neighboring stars \dotfill &$0.199\pm0.030$\\
\smallskip\\\multicolumn{2}{l}{Telescope Parameters:}&TRES\smallskip\\
~~~~$\gamma_{\rm rel}$\dotfill &Relative RV Offset (m/s)\dotfill &$2600^{+64}_{-65}$\\
~~~~$\sigma_J$\dotfill &RV Jitter (m/s)\dotfill &$167^{+94}_{-51}$\\
~~~~$\sigma_J^2$\dotfill &RV Jitter Variance \dotfill &$28000^{+41000}_{-15000}$\\
\smallskip\\\multicolumn{2}{l}{Transit Parameters:}&TESS UT 2019-01-12 (TESS)\\
~~~~$\sigma^{2}$\dotfill &Added Variance \dotfill &$0.00000019^{+0.00000020}_{-0.00000017}$\\
~~~~$F_0$\dotfill &Baseline flux \dotfill &$1.00015\pm0.00010$\smallskip\\
\multicolumn{2}{l}{}&TESS UT 2019-02-22 (TESS)\\
~~~~$\sigma^{2}$\dotfill &Added Variance \dotfill &$0.00000014^{+0.00000018}_{-0.00000015}$\\
~~~~$F_0$\dotfill &Baseline flux \dotfill &$1.000173\pm0.000096$\smallskip\\
\multicolumn{2}{l}{}&TESS UT 2021-01-20 (TESS)\\
~~~~$\sigma^{2}$\dotfill &Added Variance \dotfill &$0.00000017^{+0.00000038}_{-0.00000035}$\\
~~~~$F_0$\dotfill &Baseline flux \dotfill &$1.00054\pm0.00011$\smallskip\\
\multicolumn{2}{l}{}&TESS UT 2023-01-19 (TESS)\\
~~~~$\sigma^{2}$\dotfill &Added Variance \dotfill &$0.0000230^{+0.0000014}_{-0.0000013}$\\
~~~~$F_0$\dotfill &Baseline flux \dotfill &$0.99994\pm0.00016$\smallskip\\
\multicolumn{2}{l}{}&LCO-CTIO UT 2023-03-13 (z')\\
~~~~$\sigma^{2}$\dotfill &Added Variance \dotfill &$0.00000369^{+0.00000060}_{-0.00000054}$\\
~~~~$F_0$\dotfill &Baseline flux \dotfill &$1.00031\pm0.00020$\\
\enddata
\end{deluxetable*}

\startlongtable
\begin{deluxetable*}{lcc}
\tablecaption{Median values and 68\% confidence interval for TOI5422. \label{tab:5422_fitresults}}
\tablehead{\colhead{~~~Parameter} & \colhead{Units} & \multicolumn{1}{c}{Values}}
\startdata
\smallskip\\\multicolumn{2}{l}{Stellar Parameters:}&\smallskip\\
~~~~$M_*$\dotfill &Mass (\msun)\dotfill &$1.051^{+0.082}_{-0.063}$\\
~~~~$R_*$\dotfill &Radius (\rsun)\dotfill &$1.480^{+0.047}_{-0.042}$\\
~~~~$R_{*,SED}$\dotfill &Radius$^{1}$ (\rsun)\dotfill &$1.482^{+0.019}_{-0.018}$\\
~~~~$L_*$\dotfill &Luminosity (\lsun)\dotfill &$2.15^{+0.14}_{-0.13}$\\
~~~~$F_{Bol}$\dotfill &Bolometric Flux (cgs)\dotfill &$0.000000000568^{+0.000000000037}_{-0.000000000033}$\\
~~~~$\rho_*$\dotfill &Density (cgs)\dotfill &$0.465^{+0.037}_{-0.049}$\\
~~~~$\log{g}$\dotfill &Surface gravity (cgs)\dotfill &$4.124^{+0.029}_{-0.037}$\\
~~~~$T_{\rm eff}$\dotfill &Effective Temperature (K)\dotfill &$5744^{+39}_{-38}$\\
~~~~$T_{\rm eff,SED}$\dotfill &Effective Temperature$^{1}$ (K)\dotfill &$5740^{+110}_{-100}$\\
~~~~$[{\rm Fe/H}]$\dotfill &Metallicity (dex)\dotfill &$-0.007\pm0.039$\\
~~~~$[{\rm Fe/H}]_{0}$\dotfill &Initial Metallicity$^{2}$ \dotfill &$0.062\pm0.045$\\
~~~~$Age$\dotfill &Age (Gyr)\dotfill &$8.2\pm2.4$\\
~~~~$A_V$\dotfill &V-band extinction (mag)\dotfill &$0.197^{+0.081}_{-0.084}$\\
~~~~$\sigma_{SED}$\dotfill &SED photometry error scaling \dotfill &$0.79^{+0.33}_{-0.20}$\\
~~~~$\varpi$\dotfill &Parallax (mas)\dotfill &$2.874\pm0.023$\\
~~~~$d$\dotfill &Distance (pc)\dotfill &$347.9\pm2.8$\\
\smallskip\\\multicolumn{2}{l}{Planetary Parameters:}&b\smallskip\\
~~~~$P$\dotfill &Period (days)\dotfill &$5.377219\pm0.000010$\\
~~~~$R_P$\dotfill &Radius (\rj)\dotfill &$0.815^{+0.031}_{-0.026}$\\
~~~~$M_P$\dotfill &Mass (\mj)\dotfill &$27.7^{+1.4}_{-1.1}$\\
~~~~$T_C$\dotfill &Time of conjunction$^{4}$ (\bjdtdb)\dotfill &$2459542.2530\pm0.0011$\\
~~~~$T_T$\dotfill &Time of minimum projected separation$^{5}$ (\bjdtdb)\dotfill &$2459542.2530\pm0.0011$\\
~~~~$T_0$\dotfill &Optimal conjunction Time$^{6}$ (\bjdtdb)\dotfill &$2459961.67615\pm0.00073$\\
~~~~$a$\dotfill &Semi-major axis (AU)\dotfill &$0.0616^{+0.0015}_{-0.0012}$\\
~~~~$i$\dotfill &Inclination (Degrees)\dotfill &$88.44^{+0.97}_{-0.86}$\\
~~~~$e$\dotfill &Eccentricity \dotfill &$0.0942^{+0.0048}_{-0.0045}$\\
~~~~$\omega_*$\dotfill &Argument of Periastron (Degrees)\dotfill &$0.6^{+6.0}_{-5.8}$\\
~~~~$e\cos{\omega_*}$\dotfill & \dotfill &$0.0937\pm0.0046$\\
~~~~$e\sin{\omega_*}$\dotfill & \dotfill &$0.0009^{+0.010}_{-0.0094}$\\
~~~~$M_P\sin i$\dotfill &Minimum mass (\mj)\dotfill &$27.7^{+1.4}_{-1.1}$\\
~~~~$M_P/M_*$\dotfill &Mass ratio \dotfill &$0.02514^{+0.00058}_{-0.00068}$\\
~~~~$T_{eq}$\dotfill &Equilibrium temperature$^{7}$ (K)\dotfill &$1355^{+24}_{-18}$\\
~~~~$\fave$\dotfill &Incident Flux (\fluxcgs)\dotfill &$0.758^{+0.055}_{-0.040}$\\
~~~~$\tau_{\rm circ}$\dotfill &Tidal circularization timescale (Gyr)\dotfill &$342^{+55}_{-65}$\\
~~~~$K$\dotfill &RV semi-amplitude (m/s)\dotfill &$3068^{+22}_{-23}$\\
~~~~$R_P/R_*$\dotfill &Radius of planet in stellar radii \dotfill &$0.05664^{+0.00067}_{-0.00064}$\\
~~~~$a/R_*$\dotfill &Semi-major axis in stellar radii \dotfill &$9.00^{+0.23}_{-0.32}$\\
~~~~$\delta$\dotfill &$\left(R_P/R_*\right)^2$ \dotfill &$0.003208^{+0.000076}_{-0.000072}$\\
~~~~$\delta_{\rm TESS}$\dotfill &Transit depth in TESS (fraction)\dotfill &$0.003738\pm0.000078$\\
~~~~$\tau$\dotfill &Ingress/egress transit duration (days)\dotfill &$0.01109^{+0.0010}_{-0.00059}$\\
~~~~$T_{14}$\dotfill &Total transit duration (days)\dotfill &$0.1948\pm0.0017$\\
~~~~$b$\dotfill &Transit Impact parameter \dotfill &$0.24^{+0.12}_{-0.15}$\\
~~~~$b_S$\dotfill &Eclipse impact parameter \dotfill &$0.24^{+0.12}_{-0.15}$\\
~~~~$\rho_P$\dotfill &Density (cgs)\dotfill &$64.2^{+5.8}_{-7.4}$\\
~~~~$logg_P$\dotfill &Surface gravity \dotfill &$5.019^{+0.027}_{-0.038}$\\
~~~~$\Theta$\dotfill &Safronov Number \dotfill &$3.97^{+0.14}_{-0.15}$\\
~~~~$T_P$\dotfill &Time of Periastron (\bjdtdb)\dotfill &$2459541.078^{+0.091}_{-0.087}$\\
~~~~$T_S$\dotfill &Time of eclipse (\bjdtdb)\dotfill &$2459539.885\pm0.016$\\
~~~~$T_A$\dotfill &Time of Ascending Node (\bjdtdb)\dotfill &$2459541.070^{+0.018}_{-0.017}$\\
~~~~$T_D$\dotfill &Time of Descending Node (\bjdtdb)\dotfill &$2459543.756^{+0.018}_{-0.020}$\\
~~~~$d/R_*$\dotfill &Separation at mid transit \dotfill &$8.89^{+0.28}_{-0.33}$\\
~~~~$P_T$\dotfill &A priori non-grazing transit prob \dotfill &$0.1061^{+0.0041}_{-0.0032}$\\
~~~~$P_{T,G}$\dotfill &A priori transit prob \dotfill &$0.1188^{+0.0046}_{-0.0036}$\\
~~~~$P_S$\dotfill &A priori non-grazing eclipse prob \dotfill &$0.1056^{+0.0041}_{-0.0026}$\\
~~~~$P_{S,G}$\dotfill &A priori eclipse prob \dotfill &$0.1182^{+0.0047}_{-0.0029}$\\
\smallskip\\\multicolumn{2}{l}{Wavelength Parameters:}&TESS\smallskip\\
~~~~$u_{1}$\dotfill &linear limb-darkening coeff \dotfill &$0.303^{+0.023}_{-0.022}$\\
~~~~$u_{2}$\dotfill &quadratic limb-darkening coeff \dotfill &$0.279\pm0.022$\\
\smallskip\\\multicolumn{2}{l}{Telescope Parameters:}&TRES\smallskip\\
~~~~$\gamma_{\rm rel}$\dotfill &Relative RV Offset (m/s)\dotfill &$2265^{+19}_{-18}$\\
~~~~$\sigma_J$\dotfill &RV Jitter (m/s)\dotfill &$42^{+30}_{-22}$\\
~~~~$\sigma_J^2$\dotfill &RV Jitter Variance \dotfill &$1800^{+3400}_{-1400}$\\
\smallskip\\\multicolumn{2}{l}{Transit Parameters:}&TESS UT 2021-09-19 (TESS)\\
~~~~$\sigma^{2}$\dotfill &Added Variance \dotfill &$0.000000122^{+0.000000080}_{-0.000000077}$\\
~~~~$F_0$\dotfill &Baseline flux \dotfill &$1.000113\pm0.000031$\smallskip\\
\multicolumn{2}{l}{}&TESS UT 2021-10-16 (TESS)\\
~~~~$\sigma^{2}$\dotfill &Added Variance \dotfill &$0.000000101^{+0.000000078}_{-0.000000075}$\\
~~~~$F_0$\dotfill &Baseline flux \dotfill &$1.000064\pm0.000031$\smallskip\\
\multicolumn{2}{l}{}&TESS UT 2021-11-12 (TESS)\\
~~~~$\sigma^{2}$\dotfill &Added Variance \dotfill &$0.000000014^{+0.000000084}_{-0.000000081}$\\
~~~~$F_0$\dotfill &Baseline flux \dotfill &$1.000086\pm0.000032$\smallskip\\
\multicolumn{2}{l}{}&TESS UT 2023-10-18 (TESS)\\
~~~~$\sigma^{2}$\dotfill &Added Variance \dotfill &$0.000002217^{+0.000000090}_{-0.000000089}$\\
~~~~$F_0$\dotfill &Baseline flux \dotfill &$1.000119\pm0.000023$\smallskip\\
\multicolumn{2}{l}{}&TESS UT 2023-11-14 (TESS)\\
~~~~$\sigma^{2}$\dotfill &Added Variance \dotfill &$0.000000181^{+0.000000067}_{-0.000000066}$\\
~~~~$F_0$\dotfill &Baseline flux \dotfill &$1.000099\pm0.000020$\\
\enddata
\tablenotetext{}{See Table 3 in \cite{Eastman2019} for a detailed description of all parameters}
\tablenotetext{1}{This value ignores the systematic error and is for reference only}
\tablenotetext{2}{The metallicity of the star at birth}
\tablenotetext{3}{Corresponds to static points in a star's evolutionary history. See \S2 in \cite{Dotter2016}.}
\tablenotetext{4}{Time of conjunction is commonly reported as the "transit time"}
\tablenotetext{5}{Time of minimum projected separation is a more correct "transit time"}
\tablenotetext{6}{Optimal time of conjunction minimizes the covariance between $T_C$ and Period}
\tablenotetext{7}{Assumes no albedo and perfect redistribution}
\end{deluxetable*}